\documentclass[%
 reprint,
 amsmath,amssymb,
prx,
]{revtex4-2}

\usepackage{graphicx}
\usepackage{dcolumn}
\usepackage{bm}
\newcommand{\bv}[1]{\boldsymbol{\mathbf{#1}}}
\DeclareMathOperator*{\argmin}{arg\,min}

\begin{document}

\preprint{APS/123-QED}

\title{What are the advantages of ghost imaging? Multiplexing for x-ray and electron imaging}

\author{Thomas J. Lane}
\email{tjlane@slac.stanford.edu}
\author{Daniel Ratner}
\email{dratner@slac.stanford.edu}
\affiliation{%
 SLAC National Accelerator Laboratory,
 2575 Sand Hill Road, Menlo Park CA 94025 USA
}%

\date{\today}

\begin{abstract}
Ghost imaging, Fourier transform spectroscopy, and the newly developed Hadamard transform crystallography are all examples of multiplexing measurement strategies. Multiplexed experiments are performed by measuring multiple points in space, time, or energy simultaneously. This contrasts to the usual method of systematically scanning single points. How do multiplexed measurements work and  when they are advantageous? Here we address these questions with a focus on applications involving x-rays or electrons. We present a quantitative framework for analyzing the expected error and radiation dose of different measurement scheme that enables comparison. We conclude that in very specific situations, multiplexing can offer improvements in resolution and signal-to-noise. If the signal has a sparse representation, these advantages become more general and dramatic, and further less radiation can be used to complete a measurement.
\end{abstract}

\maketitle


\section{Introduction}

There has been recent interest in the use of \emph{multiplex} sensing techniques in applications at the atomic scale \cite{Yorke:2014bda, Yu:2016hua, Pelliccia:2016dsa, Zhang:2018fl, Pelliccia:2018ff, Li:2018ega, Kim:2018vu, Kingston:2018fl}. Traditionally, measurements using x-rays or electrons systematically interrogate a system at a single point in space, time, or photon/electron energy at once before moving on to the next measurement. In contrast, multiplex methods measure a combination of space/time/energy points simultaneously. Many repeated multiplexed measurements can convey the same information as a systematic scan if the points combined are chosen wisely. In some cases a multiplex scheme has clear advantages, either fundamental or practical. In other cases, systematic scanning is preferable. The purpose of this paper is to provide a framework for reasoning about which is best in the context of x-ray and electron science.

Through a very general analysis, we are able to conclude that for signals linear in the x-ray or electron intensity:
\begin{enumerate}
    \item Multiplex scanning enables the experimenter to swap the resolution -- in time, space, or energy -- of a final detector for the resolution with which the input beam can be modulated or measured.
    \item We confirm multiplexing can help overcome specific kinds of noise, a well-known result known as Felgett's advantage. We show, however, this advantage offers no possibility to acquire more information for a given radiation dose as compared to systematic scanning.
    \item Multiplex measurements should yield significant improvements in experiment time, radiation dose used, and signal-to-noise if the sample is known to be sparse in some basis. Multiplexing methods can lead to very general applications of the theory of compressed sensing, which represents an opportunity for imaging at the atomic scale.
\end{enumerate}
Each of these possible advantages will be discussed in detail, with clear applicability conditions. To aid readability, nearly all mathematics has been relegated to the appendices and only results are presented.

To build some intuition for how multiplex sensing techniques work, consider acquiring an x-ray fluorescence image of a sample. We could go about this in the usual way: by rastering a focused x-ray probe over the sample, thereby  illuminating one ``pixel'' at a time, and recording a spectrum at each position. At the end of the experiment, the fluorescence image is simply the set of all acquired spectra, with each spectrum assigned to a spatial location on the image.

Alternatively, we could de-focus the x-ray beam and modulate the transverse intensity distribution with a random mask. This mask might illuminate half of the pixels from our previous raster scan, chosen at random. Then, a measurement acquires the summed spectra from all the illuminated pixels, which by itself contains no spatial information. Imagine, however, we fabricate a large number of such masks and repeat the experiment many times. Every time a specific pixel $i$ is illuminated, it will contribute the $k$ points in its spectrum $\bv{x}_i$ to the spectral sum; for mask $m$, let's write that measured sum
\[
\bv{r}_m = \sum_{i} a_{mi} \cdot \bv{x}_i
\]
where $a_{mi}$ is 1 if the pixel $i$ is illuminated by the mask $m$, and 0 otherwise. For many repeated measurements this can be nicely written in matrix notation
\[
R = A X \,.
\]
This form hints that if we build our mask set $A$ in a smart way and take enough measurements, we may be able to easily solve for $X$, the same per-pixel spectra as we acquired in our raster scan.

\begin{figure}
    \centering
    \includegraphics[width=8.6cm]{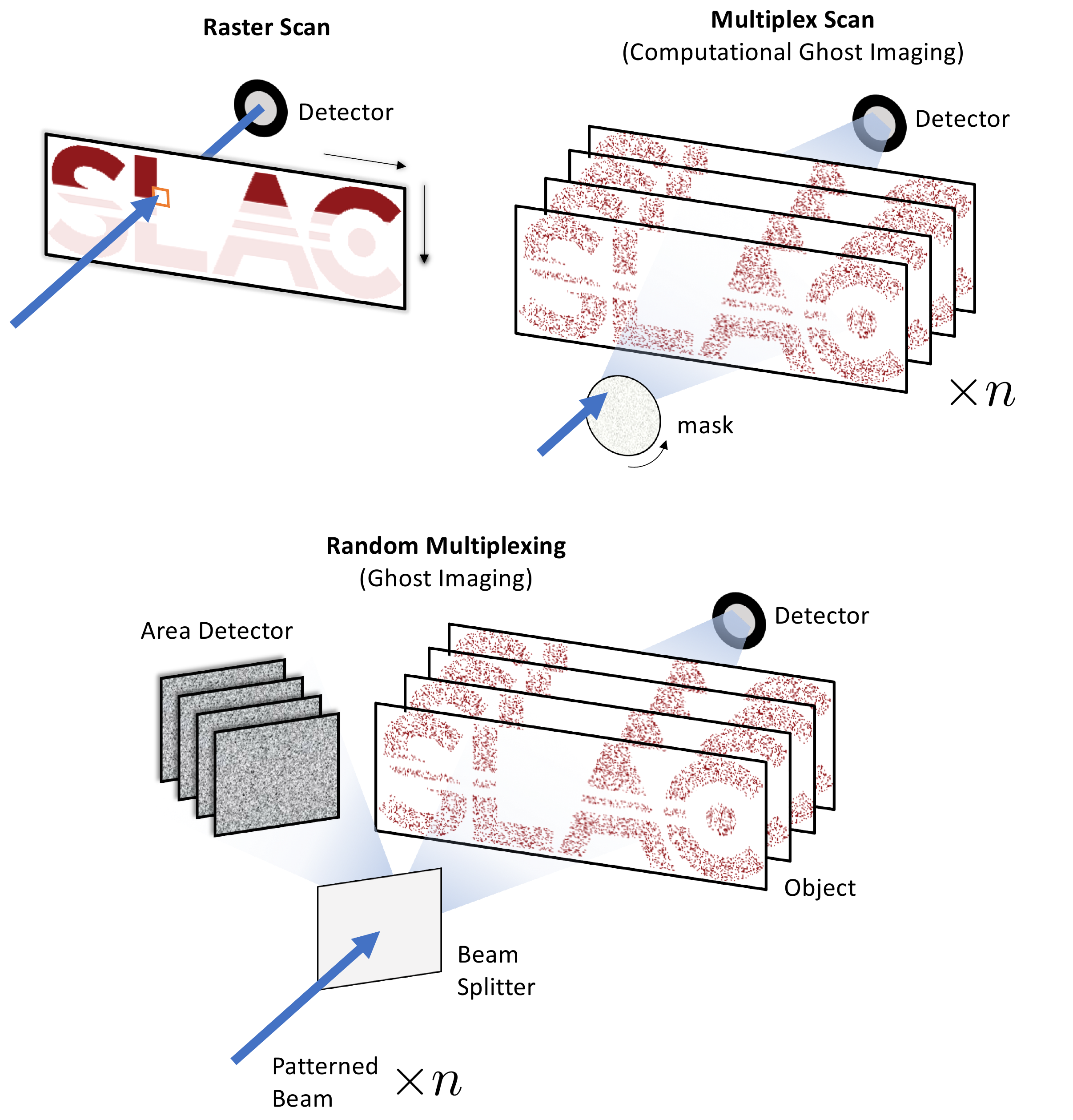}
    \caption{Three different sampling schemes for a single imaging task. Top left: in a raster scan, small regions of the object are illuminated and recorded one at a time. Top right: in a multiplex scan, many selected regions are illuminated and recorded at the same time, and their sum recorded. The patterns of illumination are chosen, so they are known. The procedure is repeated many times and the final object reconstructed mathematically. Bottom: a second example of a multiplex scan, ghost imaging. In a ghost imaging setup, a randomly patterned illumination is used and the sum recorded. The patterning, though random and uncontrolled, is measured via a diagnostic, typically a beam splitter and area detector. The reconstruction procedure is exactly the same as for the multiplex scan.}
    \label{fig:sampling}
\end{figure}

In the example presented, we specifically designed the masks and therefore knew exactly how to write our matrix $A$, with 1s and 0s for illuminated or shadowed pixels for each mask. The scheme would work just as well, however, if we had random masks we did not choose specifically, so long as we were able to measure and record which pixels were illuminated. This gives us ability to use the randomness inherent in the experiment to our advantage -- so long as we can measure that random patterning to high resolution. Figure \ref{fig:sampling} shows an illustration of all three methods discussed for acquiring a signal: raster scanning, multiplexing with known patterning, and multiplexing using measured patterning. The example envisioned is spatial in nature. The same procedure works just as well, however, in the time or energy (spectral) domains.

When is multiplexing advantageous? In this paper we provide a framework for analyzing this question, then discuss three specific cases where multiplex sensing may be useful. Our focus is on applications involving the use of x-rays and electrons, where measurements are in many ways expensive: they are often flux-limited, cause irreversible damage to the sample, can be difficult to manipulate, and require time at costly facilities. Such practical concerns will color the discussion. 

An important note on the use of prior information in imaging experiments is necessary. In general, if one knows something about the signal about to be acquired, for example if it is smooth or always-positive, that fact can be used to improve the measurement process. These priors can be implemented equally well in any sampling scheme, whether rastering or multiplexed. There is one powerful prior, however, for which this equivalence is known to not be true. If the signal is sparse in some representation (\textit{i.e.}~compressible) then multiplex sensing can offer serious advantages over raster scanning \cite{Candes:2008hb, Foucart:wp}. Because of this unique feature, sparsity will be the only prior discussed in this paper. We will first consider sampling schemes that do not make use of sparsity, then show how if the signal is known to be sparse, that information can be readily incorporated into the analysis of different imaging schemes.

Our analysis builds on progress from a variety of fields. One specific implementation of multiplex sensing is classical ``ghost imaging'' (Fig.~\ref{fig:sampling}), which refers to a class of experiments where a randomly patterned wavefront is split into two branches \cite{Padgett:2017co}. One branch measures the total transmission through a sample of interest, called the bucket signal. The other branch images the wavefront. A moment of thought reveals this process to be a case of spatial multiplexing, identical to the imaging fluorescence spectroscopy scan discussed previously, only using random beam patterning instead of known masks. In the ghost imaging community, multiplexing with known masks is conversely known as ``computational ghost imaging'' \cite{Shapiro:2008dt}. Similarly, Fourier transform spectroscopy is also a multiplexing scheme, where many wavelengths are measured simultaneously per measurement \cite{Sloane:1979tg, Harwit:1979wf}. Our presentation generalizes these examples and many others into a single framework. Our treatment is completely classical in nature, ignoring any information gain that might be achieved via performing measurements with entangled particles. Entanglement has been focus of specific studies in ghost imaging -- for perspective see \cite{Padgett:2017co} -- but is not heavily used in current applications. 

While the idea to use multiplexing in x-ray and electron experiments has only recently started to recieve significant attention, demonstrations at synchrotrons \cite{Yorke:2014bda, Pelliccia:2016dsa, Yu:2016hua}, FELs \cite{Kim:2018vu}, and electron sources \cite{Li:2018ega} have already been conducted, paving the way for multiplexing to start to impact applied x-ray and electron science.

\section{Framework}

Consider a measurement process where we perform $m$ measurements with $n$ unknown \emph{channels}, for example the pixels in an image or amplitudes for given frequencies in a spectrometer. The channel values are the unknowns we wish to infer. We measure the weighted sum of all channels; weights may be zero. Crucially, we assume the system response is linear in all aspects, including as a function of beam intensity and the number/character of the channels. Further, we consider the general case where each channel's output can be described by a $k$-length vector, for example $k=3$ if a pixel gives a simultaneous scalar output for red, green, and blue when measured, and these colors can be recorded separately. Finally, we assume some additive noise in each measurement. 

Our goal is choose a weighting scheme that will allow us to infer each individual channel's $k$ dimensional response. We write this symbolically as follows. Let $R$ be the $m \times k$ matrix of recorded outcomes, $A$ be the $m \times n$ sensing matrix, and $X$ the $n \times k$ signal of interest. Then we consider a measurement process in the presence of uncorrelated, additive noise $\epsilon$ (an $m \times k$ matrix, but where all $k$ pixels are assumed to have equivalent noise) with finite variance $\sigma^2$,
\begin{equation}
\label{eq:master}
R = AX + \epsilon \,.
\end{equation}
The details of this additive noise will be discussed in a moment.\footnote{We have not considered the case where there are errors in the record of the sensing matrix $A$. Such errors are possible when using random $A$ matrices generated naturally as part of the experiment and then measured by some diagnostic, for example as is done in ghost imaging (see Fig.~\ref{fig:sampling}). Sensing matrix errors make our estimates of $X$ too small, an effect known as \textit{regression dilution}. Regression dilution has been studied in detail in the field of statistics and we refer the interested reader to \cite{Draper:1998ab}.}
We will consider a solution to (\ref{eq:master}) to be the estimate $\hat{X}$ that minimizes the error function
\begin{equation}
\label{eq:optimize}
\hat{X} = \argmin_{X} || R - A X ||_2^2 \,,
\end{equation}
appropriate for Gaussian errors on the model predictions. For some experiments, other error models may be more appropriate, \textit{e.g.}~Poisson errors in low-flux situations. A brief comment on extending the results here to those more specific cases is given in Appendix \ref{sec:generalizedmodels}. In the rest of the text we consider Gaussian errors. 

The solution to (\ref{eq:optimize}) is given by the ordinary least squares solution,
\begin{equation}
\label{eq:ols-solution}
\hat{X} = (A^T A)^{-1} A^T R \,,
\end{equation}
from which we can compute the total expected error \emph{per channel},
\begin{equation}
\label{eq:error}
\mathcal{E} = \sigma \sqrt{ n^{-1} 
\Big\langle
  \mathrm{Tr} \big\{ (A^T A)^{-1} \big\}
\Big\rangle_A
} \,.
\end{equation}
While the ordinary least squares solutions are well known, for completeness in Appendix \ref{app:ols_proof} we compactly prove (\ref{eq:ols-solution}) and (\ref{eq:error}). We are motivated to do so by the fact that in the field of ghost imaging it is still common to use sub-optimal estimators, even though least squares has been employed previously \cite{Katz:2009fv, Zhang:2014go, Jiang:19}. The properties and performance of the traditionally used ghost imaging estimator is discussed in Appendix \ref{sec:traditional_gi} and shown numerically in Fig.~\ref{fig:simulation}.

Equation (\ref{eq:error}) naturally separates the error contributions from the noise, given by $\sigma$, and the stability of the inversion of $A$ represented by the trace term $\mathrm{Tr} (A^T A)^{-1}$. As we will see, both depend on the choice of $A$, and are therefore determined by the experimental design. The noise factor, $\sigma$, depends also on the noise characteristics of the detection method, which turns out to be important for the comparison between different sampling schemes.

\subsection{Noise Model and Transferred Noise}

While in general the presented equations are applicable in any case of additive noise, in real experiments the noise can often be broken into three standard types:
\begin{enumerate}
    \item Quantum Poisson noise
    \item Per-channel (\textit{e.g.}~per-pixel) Gaussian noise, often caused by the electronics in a detector
    \item Total-detector Gaussian readout noise that is incurred once per measurement.
\end{enumerate}
Considering these three types of noise with variances $\sigma_p^2$, $\sigma_x^2$ and $\sigma_m^2$ respectively, we can approximate the total additive noise $\sigma$ as a function of the magnitudes of the individual contributions and the experimental design,
\[
\sigma = \sqrt{ \left\langle  
  \sigma_p^2 \sum_{i=1}^n a_{mi} + \sigma_x^2 \sum_{i=1}^n a_{mi}^2 
+ \sigma_m^2 
 \right\rangle_A }
\]
where we have assumed each column $\bv{a}_m$ of the matrix $A$ is identically and independently distributed (Appendix \ref{sec:additive_error}). We have also assumed that we known nothing about the structure of the sample \textit{a priori}, and therefore assume a uniform signal intensity purpose of estimating Poisson noise (Appendix \ref{sec:poisson_error}).

Typically, one source of noise will be dominant in the system. In that case, the other sources will be negligible and we can approximate,
\begin{align}
\label{eq:error_cases}
    \sigma \approx
    \begin{cases}
      \sigma_p \sqrt{ n \langle a_{mi} \rangle} \ \ &\mathrm{Poisson} \\
      \sigma_x \sqrt{ n \langle a_{mi}^2 \rangle} \ &\mathrm{Channel \ Gaussian} \\
      \sigma_m  \ &\mathrm{Measurement \ Gaussian} \\
    \end{cases}
\end{align}
Note that for the per-channel cases, the noise scales as $\sqrt{n}$, the square root of the number of channels. This should be intuitive, as adding an additional channel brings with it that channel's noise contribution. It also brings that channel's contribution to the summed signal, so these factors cancel with the $n^{-1/2}$ factor in equation (\ref{eq:error}) to make the overall error $\mathcal{E}$ independent of the number of channels.

More interesting is the case where the error is proportional only to the number of measurements. This occurs for example if there is some significant constant readout noise for a detector -- then, the number of channels doesn't matter, only the number of readouts (measurements). In this case, the overall error is reduced by increasing the number of channels and scales as $\mathcal{E} \sim n^{-1/2}$. The advantage comes from each channel contributing additional intensity, and that intensity boosts the signal higher and higher above the constant noise floor. This is well known in the field of spectroscopy, where the $n^{-1/2}$ reduction of the noise as a function of multiple channels is known as the Felgett advantage, one motivation behind Fourier transform and Hadamard spectroscopies. See \cite{Harwit:1979wf} for a comprehensive study of noise in multiplex spectroscopy.

The question is if this kind of noise is applicable in modern x-ray and electron imaging setups. Current x-ray detectors are able to routinely count individual photons with high quantum efficiency, which effectively eliminates readout noise of this sort \cite{Forster:2019kz}. Electron detectors have also progressed into the quantum counting regime recently \cite{Kuhlbrandt:2014ho}. The major source of noise in modern setups, therefore, is quantum (Poisson) in nature, and occurs on a per-channel basis. The traditional Felgett advantage does not apply in such situations, and no signal-to-noise advantage is achieved by multiplexing.

\subsection{Inversion Stability: Trace Term}

Equation (\ref{eq:ols-solution}) assumes that $A^T A$ is non-singular and can be inverted, which in general is not true, especially if $A$ is random. In such cases approximate -- but often very good -- solutions may be found by solving Eq.~(\ref{eq:optimize}) via a matrix factorization or a minimization algorithm. Reasonable solutions will only be obtained if $A^T A$ is high rank, and good design of a sampling scheme should ensure that $A^T A$ is non-singular with high probability in the limit of a reasonable number of samples. The schemes we discuss here are all of this variety. In the case of random sensing matrices, however, $A^T A$ may only be strictly non-singular in the limit of a large number of random samples, and errors computed using (\ref{eq:error}) will be lower bounds.

\subsection{Radiation Dose}

In this model, the total dose on the sample $\mathcal{D}$ for a set of measurements is given by the sum of all the elements of $A$,
\begin{equation}
\label{eq:dose}
\mathcal{D} = \left\langle
  \sum_{ij} |A_{ij}|
\right\rangle_A \,.
\end{equation}
We assume the damage is linear in total dose, and do not consider cases where the damage mechanism changes as a function of fluence, \textit{i.e.}~changes in dose-intensity per unit time or area.
\begin{figure*}
    \centering
    \includegraphics[width=17.6cm]{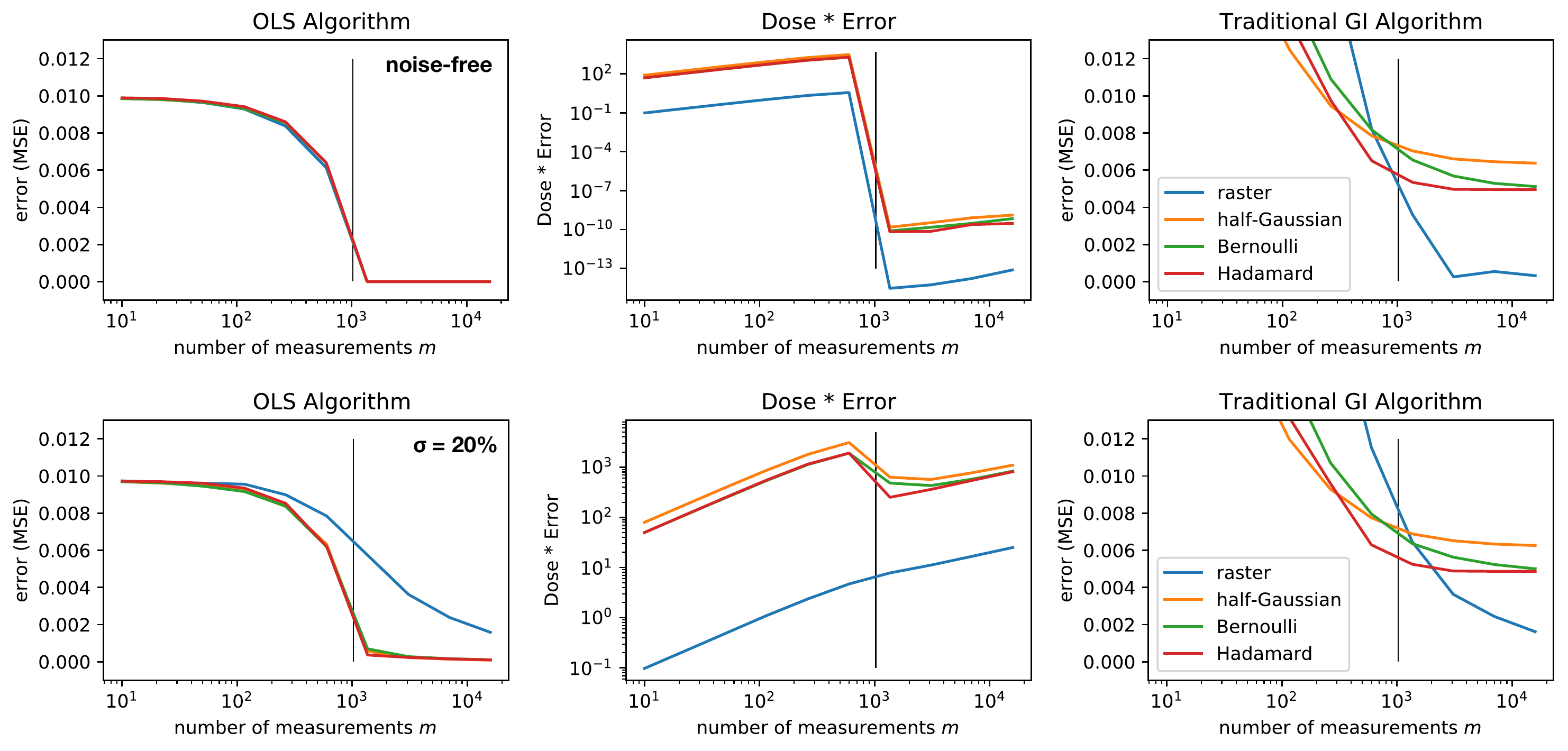}
    \caption{Simulations showing the performance of different sampling schemes: raster, Bernoulli, half-Gaussian, and Hadamard. A 1023-length waveform with iid random samples from $\mathcal{N}(0,1)$ smoothed using a size-15 median filter was used as the signal of interest. This signal was reconstructed using either the ordinary least squares estimator (Eq.~\ref{eq:ols-solution}) [left], for which the dose-error product is shown [middle], or the traditional ghost imaging correlation algorithm (Eq.~\ref{eq:traditional_gi}) [right]. Top: in the absence of noise, Bottom: in the presence of 20\% ($\sigma_m = 0.2$) per-measurement additive Gaussian noise. Each simulation was repeated 10 times with new random values for the signal, sensing matrices, and noise, with average results shown. Black line shows $m = n$. Observations: (1) the dose and error predictions of Table \ref{tab:comparison} are reproduced, (2) the Felgett advantage is clearly visible in the simulation containing noise [bottom left], (3) the performance of the traditional ghost imaging algorithm is inferior to the OLS solution as predicted by theory (Appendix \ref{sec:traditional_gi}).}
    \label{fig:simulation}
\end{figure*}

\section{The Analysis of Some Examples}

We consider four sampling schemes for a demonstration comparison: a raster setup, two random multiplexed schemes, and one deterministic multiplexing scheme. For each, we compute the error of a recovered signal as a function of the number of measurements performed and radiation dose incurred. Results are summarized in Table \ref{tab:comparison}.

\subsection{Raster Scans}

In a raster scan, each pixel is sequentially illuminated with intensity $a$ so that $A \equiv aI$. Usually the number of measurements equals the number of channels ($m = n$). In the presence of noise, it might be productive to repeat measurements. In that case $A$ once again becomes an $m \times n$ matrix where rows of the identity matrix start to repeat. 

The solution to the reconstruction problem (Eq.~\ref{eq:ols-solution}) is simply $\hat{X} = R/a$, reflecting the intuition that in raster scanning no reconstruction is necessary. The radiation dose of the experiment is $am$ while the expected error is $(\sigma \sqrt{n}) / ( a \sqrt{m} )$, showing how increasing the beam intensity overcomes background noise at the cost of possibly damaging the sample.

\subsection{Multiplex Scans}

Any non-raster scan is a multiplex scan. To highlight the differences, we will discuss situations where the sensing matrix $A$ is fairly dense, \textit{i.e.}~about half or more of the elements are non-negligible. Consider the following three concrete multiplex scans:
\begin{enumerate}
    \item Half Gaussian. $a_{ij} \sim |\mathcal{N} (0, a^2)|$. Each element distributed independently and identically (iid).
    \item Bernoulli with $p=1/2$. $a_{ij} = a$ with probability $1/2$ and $0$ otherwise, iid.
    \item Hadamard. The Hadamard sequence used in spectroscopy involves constructing a specific $n=2^z-1$ length sequence of $0$s and $1$s (with $z$ a positive integer) that forms an $n \times n$ circulant matrix $S_n$. This matrix is used as a sensing matrix.\footnote{These matrices are closely related to, but are not exactly, the traditionally defined Hadamard matrices. They are formed by generating a Hadamard matrix of order $n+1$, truncating the first row and column -- which contain only 1s -- and performing the substitions $1 \to 0$ and $-1 \to +1$.} The sequence is deterministic and the circulant property can simplify the experimental implementation. For our discussion, the unframiliar reader need only know that $S_n$ is a deterministic matrix where approximately half of the elements are $1$, the rest are $0$, and the inverse of $S_n$ is given by $ 2(n+1)^{-1} \cdot (2S_n^T - J_n) \neq S^T$. By $J_n$ we mean an $n \times n$ matrix with all elements $1$ \cite{Sloane:1979tg, Harwit:1979wf}.
\end{enumerate}
Note the first two cases are random -- continuous and binary, respectively. The last is binary and deterministic.

\subsection{Comparison}

A quantitative comparison between these four sampling schemes is presented in Table \ref{tab:comparison}, containing analytical results, and Figure \ref{fig:simulation} which contains numerical results. Some important observations are immediately noticeable:
\begin{enumerate}
    \item In cases where radiation damage is a concern, rastering always offers the best error-to-damage tradeoff. The increased damage for adding multiplexing channels is proportional to the number of channels $n$, while the error reduction offered by the Felgett advantage scales as at most $\sqrt{n}$, and at worst negligible, see Eq.~(\ref{eq:error_cases}).
    \item Hadamard sampling can be considered optimal in terms of maximizing Felgett's advantage \cite{Harwit:1979wf}. It is interesting, however, that the performance of a random Bernoulli sampling is nearly identical to the carefully designed Hadamard one.
    \item Our results highlight that the performance of all schemes are not the same. The half Gaussian sampling scheme has a higher error for a given dose when compared to the Hadamard or Bernoulli methods, for instance.
\end{enumerate}
\renewcommand{\arraystretch}{1.3}
\begin{table*}[t]
    \begin{tabular*}{\textwidth}{l| @{\extracolsep{\fill}} rrr}
        \hline
        ~         & Dose ($m$ meas.) & Error (per channel, $m$ meas.) & Dose-Error Product  \\ 
        \hline \hline
        Raster    & $a m$                       
          & $\sigma / (a \sqrt{m n}) $  
          & $ \sigma \sqrt{m/n} $ \\ 
        Half Gauss  & $\sqrt{{2 / \pi}} \cdot a m n $ 
          & $1.66 \cdot \sigma / (a \sqrt{m})$ 
          & $1.32 \cdot \sigma n \sqrt{m}$ \\ 
        Bernoulli & $(1/2) \cdot a m n$                     
          & $2 \cdot \sigma / (a \sqrt{m})$ 
          & $\sigma n \sqrt{m}$ \\ 
        Hadamard$^\ddag$  & $(1/2) \cdot a m n$ 
          & $2 \cdot \sigma / (a \sqrt{m})$ 
          & $\sigma n \sqrt{m}$ \\
        \hline
    \end{tabular*}
    \caption{Quantitative comparison between the sampling schemes discussed, produced using equations (\ref{eq:error}) and (\ref{eq:dose}). Recall $m$ is the number of measurements, $n$ is the number of channels (i.e. unknowns), $a$ is the average dose per channel, $\sigma$ is the standard deviation of the additive Gaussian noise per measurement. The factor $1.66$ in the half Gaussian distribution approximates $\left( 1 - 2/ \pi \right)^{-1/2}$. Note that all these values assume $A^T A$ is not singular, which can only occur when $m \geq n$. $^\ddag$Hadamard error estimate derived in \cite{Harwit:1979wf}.}
    \label{tab:comparison}
\end{table*}

\section{Experimental Opportunities for Multiplex Sensing}

The presented framework gives us the tools for assessing the fundamental capabilities and costs associated with a particular sampling scheme. Often, however, the benefits of multiplexing are practical. The ability to use flexible illumination in either space or time can alleviate engineering challenges, making new experiments possible. Next, we discuss situations where multiplexing can give performance advantages due to either practical or fundamental concerns.

\subsection{Resolution Improvement}

In standard, raster-style imaging, the resolution of the measurement is limited by either the probe or detector resolution. For instance, in the x-ray fluorescence imaging application discussed in the introduction, the final resolution of the hyperspectral image obtained will be limited by the x-ray focus size.

Multiplexing enables one to trade the resolution of the probe or detection system for the resolution at which the probe can be patterned or measured. This trade may make it possible to collect data at higher resolution. Alternatively, it may be a simpler or cheaper option than increasing the probe resolution.

As an example, we recently suggested along with our colleagues that by measuring the random temporal fluctuations inherent in the power of an XFEL pulse, pump-probe experiments with time resolution faster than the pulse duration would be possible. A typical XFEL pulse has a duration on the order of 10-100 fs, but its power over time can be readily measured to $\sim 1$ fs resolution using well established diagnostics such as the LCLS's XTCAV \cite{Behrens:2014ib}. This should enable x-ray pump-probe measurements to be conducted down to 1 fs resolution. The alternative of decreasing the pump and probe durations -- preparing 1 fs x-ray pulses -- is possible, but challenging, and requires new hardware to implement \cite{MacArthur:IPAC2017-WEPAB118, Hartmann:2018dq, Serkez:2018jx}. Therefore for some experiments time-domain multiplexing should offer an attractive alternative.

This highlights an interesting aspect of multiplex sensing. In cases where an experimental setup naturally results in random fluctuations in the measurement, it may be possible to use that randomness rather than fight it.

\subsection{Felgett's Advantage: More Signal per Measurement}

Multiplex scans have multiple channels illuminated at once. Thus, for a system that has a fixed flux-per-channel, multiplexed illumination offers a higher overall flux on the sample. In experiments with per-measurement noise, this enables one to achieve a specific signal-to-noise ratio with fewer measurements. Given the large investments in making brighter x-ray and electron sources and the expense associated with operating those sources, the ability to use more of the incident flux is attractive. 

It should be emphasized that the experiment time is reduced precisely because the sample is exposed to more incident radiation. This does not necessarily translate into more information acquired for a given radiation dose. Moreover, the multiplex advantage only exists with per-measurement noise (such as $\sigma_m$ in Eq.~\ref{eq:error_cases}), and there is no gain when using photon-counting detectors.

\subsection{Signal-to-Noise per Dose}

Can multiplex sensing enable one to obtain more information for a given radiation dose? Many atomic resolution imaging tasks are radiation-damage limited, and the possibility to use multiplex sensing to overcome that limit has been discussed in recent literature \cite{Pelliccia:2016dsa, Zhang:2018fl, Kim:2018vu, Pelliccia:2018ff}. This discussion has revolved around ghost imaging setups, where it has been suggested that damage might be avoided by decreasing the flux on the object branch and increasing the flux into the reference branch. Since the reference branch radiation does not interact with the object, then it follows damage will be reduced.

This argument needs to consider errors in the measurements performed on both branches. The analysis presented so far corresponds to the case of an infinitely bright reference branch with no error on the matrix $A$. In the case where such error is significant, one expects regression dilution \cite{Draper:1998ab} and our work only provides a lower bound error estimate. Here, we consider only the limiting case, where the reference branch is made so bright that the error on the measurement of $A$ is negligible. 

Now we only need to consider error on the object branch (as written in Eq.~\ref{eq:error}). Even with an ideal detector, there will at least be Poisson noise due to the fact we image with quanta (photons or electrons). 

Even in this best case scenario, our analysis shows there is no fundamental way classical ghost imaging -- or any multiplexing scheme -- can alleviate radiation damage (in the absence of priors, see section \ref{sec:cs}). Multiplexing with $n$ channels, corresponding to pixels on the reference branch, reduces the error by at most a factor of $\sqrt{n}$ (Eq.~\ref{eq:error_cases}) which occurs in the presence of per-measurement noise. From a signal-to-noise perspective, this is equivalent to reducing the number of measurements, and therefore the radiation dose, by a factor of $n$ (since error scales with the square root of the number of measurements, $\mathcal{E} \sim m^{-1/2}$). The $n$ additional channels, however, increase the dose by a factor of $n$ (Eq.~\ref{eq:dose}). Thus, the \emph{additional} dose used to obtain Felgett's advantage exactly offsets the dose reduction possible by conducting fewer measurements (see Table \ref{tab:comparison}). In the best case scenario, multiplexing does not improve the error-to-dose ratio.

In the case of a perfect detector with only quantum noise, the scaling is even worse -- multiplexing does not decrease the error at all, but still incurs the dose penalty. Alone, multiplexing cannot limit radiation damage, only make it worse.

We can understand this intuitively. Consider the limiting case in a ghost imaging setup, where for each ghost image exposure the object branch is so weak that exactly a single photon travels along that branch (zero photons would provide no information, as the bucket would always read zero). That photon is either absorbed by the object or transmitted and detected. An infinitely bright reference branch provides a probability distribution for where the photon interacted with the sample. What is the most informative possible distribution? Clearly, a delta function -- telling us \emph{exactly} where the imaging photon went -- would be most useful. This corresponds precisely to a raster scan, where we send the photon in a pre-determined and known direction and measure the result. Ghost imaging imparts at most an equal amount of information per object-branch photon as a raster scan, and in general less.

\begin{figure}
    \centering
    \includegraphics[width=8.6cm]{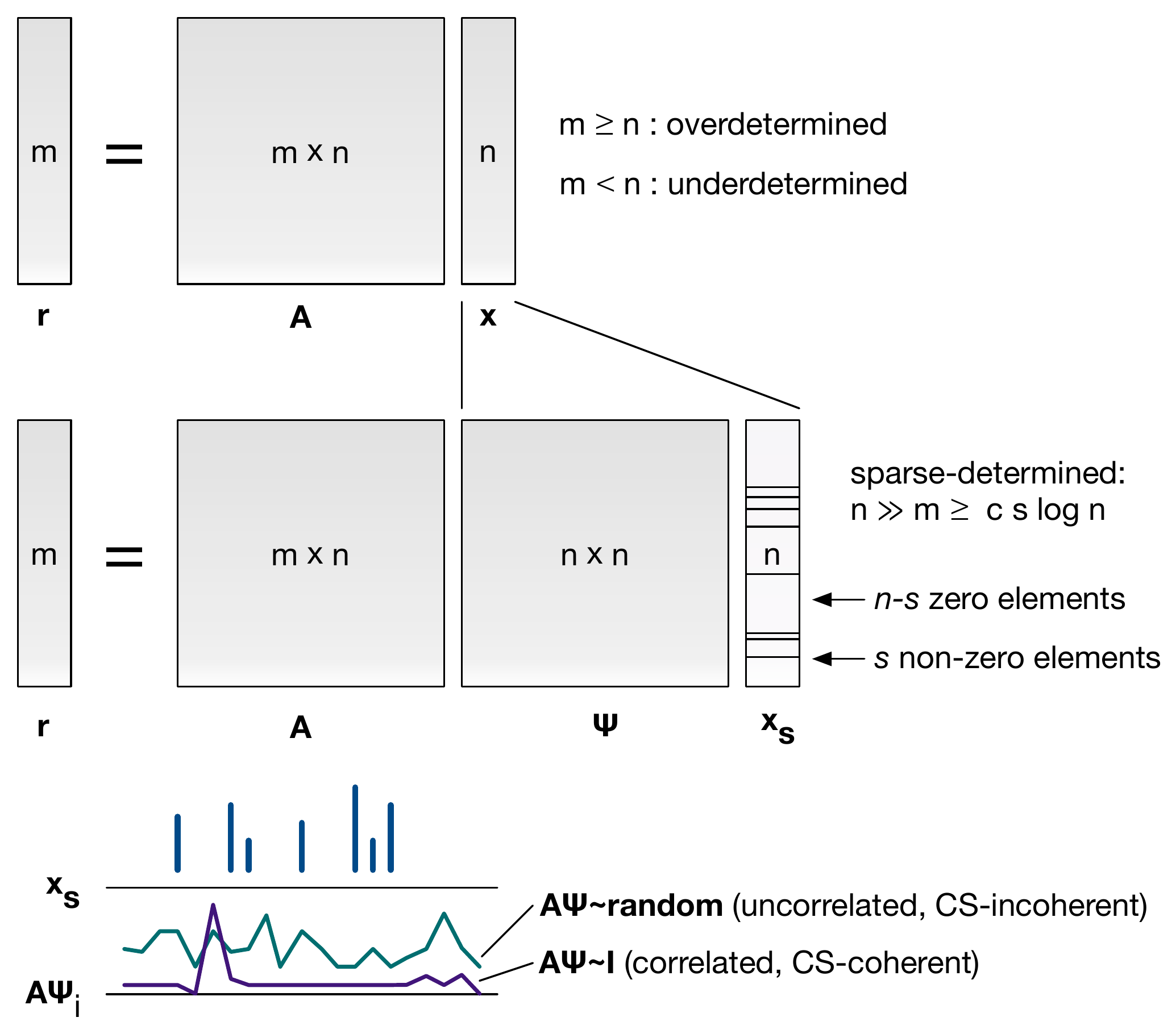}
    \caption{Compressed sensing uses sparsity to enable inference with fewer samples than unknowns. Top: when solving linear systems, if the number of unknown variables $n$ is greater than the number of known measurements $m$, then the problem is underdetermined and no unique solution exists. Middle: if, however, the signal is known to be sparse in some basis $\Psi$, then the number of knowns is effectively reduced and the problem may become solvable. It is essential, however, that the majority of the measurements provide information about the non-sparse elements of the sparse signal. Bottom: if the matrices $A$ and $\Psi$ are CS-incoherent, then this is ensured. If, however, $A$ and $\Psi$ are highly correlated, then there is a high chance any measurement only informs on zero components of the sparse signal and provides no information. The required number of samples in this case is expected to return to order $n$, as this number of samples are required to ensure all components of the sparse signal are measured.}
    \label{fig:cs}
\end{figure}

\section{Compressed Sensing}
\label{sec:cs}
Throughout our discussion we have assumed that no prior information is used in the signal reconstruction process. There is, however, one important situation where prior information about the structure of the signal can -- in combination with multiplex scanning -- dramatically improve the imaging process.

If the signal $X$ to be measured is sparse in some basis, then techniques of compressed sensing can be applied \cite{Candes:2008hb, Foucart:wp}. Here we quickly review compressed sensing theory, in order to show how multiplexed measurements can enable a general application of compressed sensing in x-ray and electron science.

Compressed sensing presumes that our signal $X$ of interest has a sparse representation. Concretely, this means that there exists some orthonormal basis $\Psi$ such that $X_s \equiv \Psi^{-1} X$ is $s$-sparse, meaning it has only $s$ non-negligible ($\approx 0$) entries. If $s \ll n$ we may be able to determine the signal $X$ of interest with order $s$ measurements instead of $n$ (Fig.~\ref{fig:cs}). If we can do so, we can obtain the signal with less time, radiation dose, and error.
%

The basis $\Psi$ can be chosen to maximize the sparsity. For example, if the signal is composed of a waveform with only a few frequencies, choosing $\Psi$ as a DFT matrix will transform the signal into the frequency domain, where it will be sparse. Further, it is absolutely possible for the signal to be sparse in the original domain of interest (for example, an image with a uniform background), in which case $\Psi = I$ is a reasonable choice.

If we knew precisely which elements in $X_s$ were sparse, the procedure to use would be straightforward. We could simply truncate $X_s$ to the non-negligible components and perform our inference as before with $s$ instead of $n$ unknowns, under the image formation model $R = A \Psi X_s$.\footnote{We are assuming a complete column of $k$ elements in $X_s$ is zero or not for the purposes of counting sparse elements. If one element is non-zero, that entire column is counted as one of the $s$ datapoints we have to infer.} If, however, we do not know which components will be sparse, it is essential that each measurement provides some new information about each non-sparse component. This is where multiplexed measurements enter the picture. To ensure that \emph{each} of our measurements are highly likely to give information about \emph{all} $s$ components, the measurement scheme $A$ must be constructed in a particular way. It turns out random multiplexing satisfies this requirement for any signal and any sparsity basis, as we will now discuss.
%

In the theory of compressed sensing, the ability of a sensing scheme to yield information about all the non-zero signal components is quantified by the \emph{coherence} $\mu$
\[
\mu( A, \Psi ) = \sqrt{n} \max_{i,j} 
\frac{\bv{a}_i \cdot \bv{\psi}_j}{ ||\bv{a}_i||_2 \cdot ||\bv{\psi}_j||_2}
\]
which is the correlation between the rows of $A$ and the basis vectors (columns) of $\Psi$. The term coherence should not be confused with the phase relationship between any waveforms, and to distinguish it we will call it CS-coherence.

\begin{figure}
    \centering
    \includegraphics[width=8.6cm]{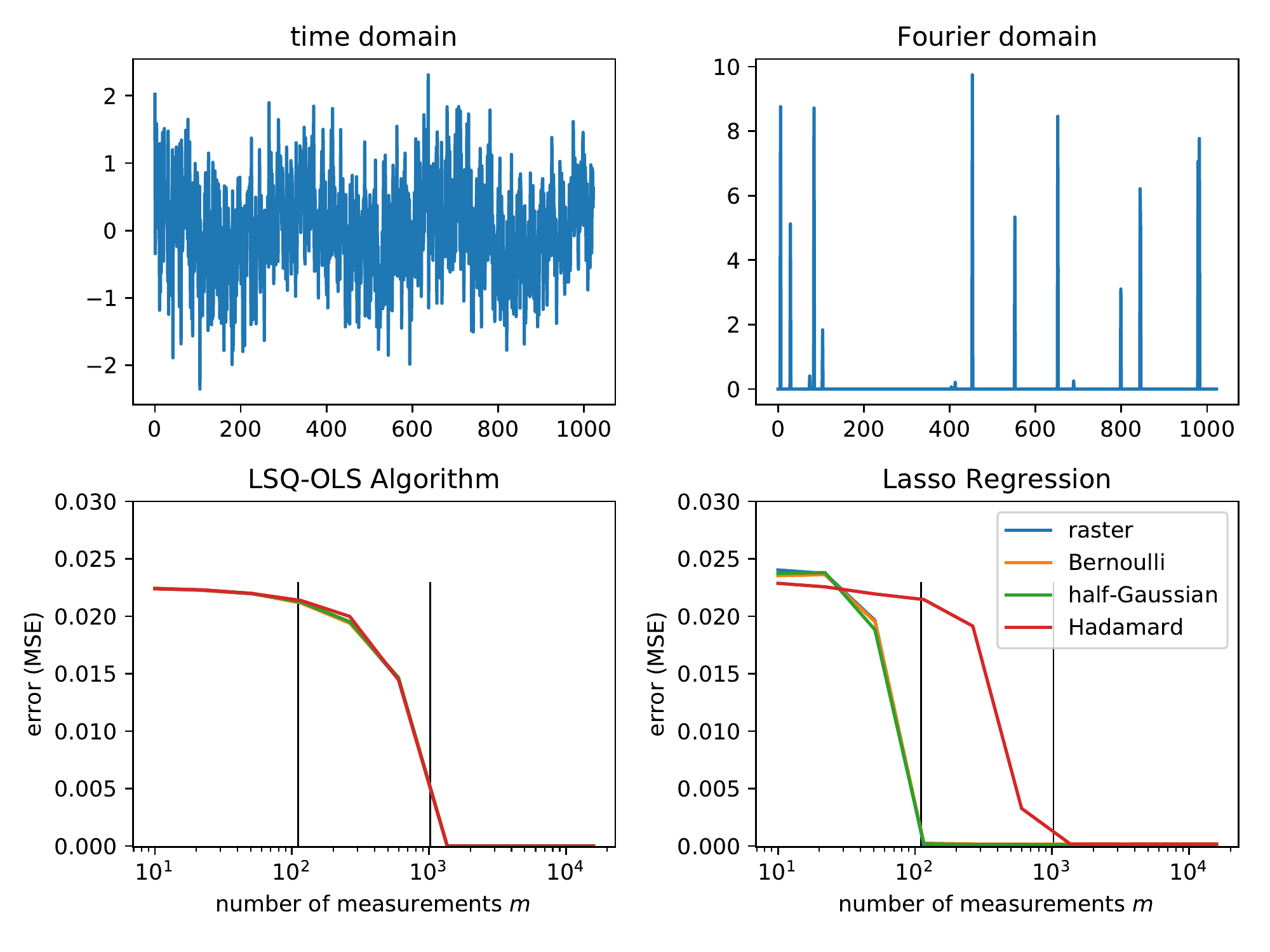}
    \caption{Example of compressed sensing using the multiplexing schemes discussed. A signal (top left) of 1023 samples was constructed by summing 16 cosine functions with random frequencies (chosen uniformly over the band limit but excluding the DC component) and positive amplitudes from $10 \times |\mathcal{N}(0,1)|$, making the signal sparse in the Fourier domain (top right). The signal was then reconstructed from a variable number of samples using either the ordinary least squares algorithm (bottom left) or the lasso algorithm with $\lambda = 10^{-2}$ for all sensing matrices, except the identity case (which has a factor of $\approx 1024 / 2$ smaller amplitude) for which $\lambda = 1 \cdot 10^{-5}$. Values of $\lambda$ were hand-tuned. Shown are the averages of 10 randomized repeats of the simulation process. On the bottom panels, the leftmost black line indicates $m = s \log n$, and the rightmost line is $m = n$. The Hadamard reconstruction requires additional samples because it is partially coherent with the Fourier basis ($\mu = 20.3$), as compared to the incoherent identity, Bernoulli, and half-Gaussian sensing bases ($\mu = 1.4, 3.4, 3.0$ respectively).}
    \label{fig:cs_simulation}
\end{figure}

Given a pair $A, \Psi$, we can replace Eq.~\ref{eq:optimize} with the lasso regression algorithm,
\[
\min_{X_s} || R - A \Psi X_s ||_2^2 + \lambda || X_s ||_1.
\]
Here, the $|| X_s ||_1$ term enforces signal sparsity and parameter $\lambda$ determined by the noise level $\epsilon$. In general this parameter is unknown, and set using cross-validation or some other estimate based on observed data. If
\[
m \geq c \, \mu^2 \cdot s \, \log n
\]
for some constant $c$ that depends on $A$, then a central result of compressed sensing theory demonstrates this program recovers the exact signal with probability $1 - O(e^{- \gamma m})$ with $\gamma > 0$ \cite{Foucart:wp}. If $s$ is small, then often $m \ll n$ measurements give an exact reconstruction. Figure \ref{fig:cs_simulation} shows this scaling is achieved using the lasso algorithm for a simple sparse reconstruction. This reduction in the number of measurements is the advantage compressed sensing offers for x-ray and electron experiments.

Consider instead trying to use a raster scan ($A=I$) to reveal a naturally sparse signal ($\Psi = I$). Here, $A \Psi = I$ and so $\mu^2 (A, \Psi) = n$, confirming the expectation that $n$ independent measurements are required to uniquely infer a signal of size $n$.  

Since we wish to presume as little as possible about the sample of interest and its sparsity structure, it is natural to ask: are there sensing matrices $A$ that are CS-incoherent with nearly all sparsity representations $\Psi$? It turns out the answer is yes, and randomly multiplexed schemes, such as the half-Gaussian and Bernoulli matrices discussed in this paper, are examples. It can be shown the random nature of such sensing matrices make it almost certain $\mu \ll n$. In other words, signals that are randomly multiplexed are guaranteed to be CS-incoherent.  This holds for any sparsity basis $\Psi$ (specifically $\mu^2 < c \log n$, and often $\mu =$ const.) \cite{Candes:2008hb, Foucart:wp}. This theory shows why multiplexing is so useful in compressed sensing applications -- random multiplexing enables us to use \emph{any} sparsity basis of our choice, thereby maximizing sparsity and the associated advantages of compressed sensing. 

As an intuitive example, consider an example where the signal of interest consists of an image with only a few localized features on blank background, but the feature locations are unknown (here, $\Psi = I$). Raster scanning for the features will be slow. In fact for $n$ pixels, $s$ of which are non-blank, we get useful information only a fraction ($s/n$) of the time we perform a raster measurement. Further, we have to scan all the pixels to ensure we don't miss any feature. Instead, consider conducting measurements where each pixel has a $50\%$ probability of being illuminated in each measurement. On average, we obtain information on fraction $s/2$ of the information-containing pixels per measurement, as compared to the average fraction $s/n$ in the raster scan. Since it is likely $n \gg 2$, this is a more efficient measurement scheme. Further, the probability of missing a given feature goes as $2^{-m}$, which means it is unlikely we miss any feature entirely, even for a modest number of measurements ($< 0.1\%$ for $m=10$).

We emphasize that by reducing the number of samples below $n$, compressed sensing is the only multiplexing approach which reduces the total dose needed to achieve a given signal-to-noise ratio.

\section{Summary}

We have presented a framework for assessing the performance of different sampling schemes. In addition to the common rastering method, where a single point in space, time or energy is measured, we have considered multiplex schemes where sets of points are measured simultaneously. In specific cases, the latter can offer advantages in experiment time, sample use, and measurement resolution: 

\begin{itemize}
\item When per-measurement readout noise is dominant, the Felgett advantage improves the signal-to-noise ratio. 
\item If the source is naturally random, multiplexed measurements using that randomness may be more convenient or yield higher resolution than producing a raster beam.
\item If the resolution at which the multiplex beam can be measured  or controlled exceeds that of a raster beam, multiplexing will improve the resolution of the final reconstruction.
\item Finally, if the acquired image is known to be sparse in some domain, compressed sensing enables experiments to be conduced with fewer measurements, improving experiment time, radiation dose, and signal-to-noise. Random multiplexing allows one to apply compressed sensing generally, to any signal, with any sparsity basis.
\end{itemize}

We have proven two facts that may be especially counterintuitive for this budding community. First, the only way multiplexing can be used to mitigate radiation damage or perform more rapid experiments is when it is combined with compressed sensing. If the signal is known to be sparse, the expected number of measurements for the multiplexing schemes scales roughly as $s \log n$, with simulations providing a more accurate final assessment.  Second, random multiplex patterns are about as good as the carefully programmed Hadamard sequence -- whichever is easier to implement experimentally should be used. Finally, we emphasize an old result that is little known in photon science; there is no Felgett advantage when Poisson noise dominates.

The advantages offered, however, are idiosyncratic to each application. When considering a multiplex sampling scheme, we recommend answering four questions first:
\begin{enumerate}
    \item Is the experimental goal limited by radiation dose, number of measurements, or resolution?
    \item Is it possible to program a multiplex scheme or measure some natural or induced randomness in the measurement to implement multiplexing?
    \item What noise is present in your signal of interest?
    \item Is your expected signal known to be sparse in some basis? If so, what is the expected sparsity?
\end{enumerate}
With the answers to these questions in hand, Eq.~\ref{eq:error} provides the final error expected for each possible experimental setup.

Large capital investments are currently being made in x-ray and electron facilities and instrumentation. It is our belief that compressed sensing offers an under-utilized opportunity to push the capabilities of these new radiation sources, making current experiments more efficient and opening new experimental possibilities. Since compressed sensing and multiplexing go hand-in-hand, this in turn means we anticipate an increased use of multiplex sensing in x-ray and electron science in the near future. Future work is necessary to develop general ways to apply compressed sensing to the wide variety of imaging currently being performed with electrons and x-rays.

Even in cases where sparsity cannot be leveraged, we believe multiplexing offers significant advantages for very specific x-ray and electron imaging experiments. Our hope is that the work presented here helps clearly identify those advantageous cases and implement them.

\subsection*{Code}

Code used to perform simulations and generate figures used in this manuscript is freely available at \url{https://gist.github.com/tjlane/35a84123e3df73cea014ff082a5ad0a8}.

\appendix

\section{OLS Proof}
\label{app:ols_proof}

For completeness, we quickly sketch proofs of the OLS solution and error analysis. We use vector notation (k=1) for simplicity, but the results are easily extended. To show the optimality of the estimator (\ref{eq:ols-solution}), we wish to minimize (\ref{eq:optimize}),
\begin{align*}
    || \bv{r} - A \bv{x} ||_2^2 =&
        \left( \bv{r} - A \bv{x} \right)^T \left( \bv{r} - A \bv{x} \right) \\
    =& \, \bv{r}^T \bv{r} - 2 \bv{x}^T A^T \bv{r} - \bv{x}^T A^T A \bv{x}
\end{align*}
differentiation with respect to $\bv{x}$ yields
\[
- 2 A^T \bv{r} + 2 A^T A \bv{x}
\]
which when set equal to $0$ shows that (\ref{eq:optimize}) is minimized when $\bv{x} = (A^T A)^{-1} A^T \bv{r}$. Note this estimator is defined for non-square $A$; for such matrices $A^{-1}$ nor $(A^{T})^{-1}$ exist. This operator is also known as the Moore-Penrose pseudoinverse and denoted $A^+ \equiv (A^T A)^{-1} A^T$, a compact notation we will use immediately. To derive the expected error of this estimator under the error model (\ref{eq:master}),
\[
\bv{r} = A \bv{x} + \bv{\epsilon}
\]
we can compute the residual ($\hat{ \bv{x} } - \bv{x}$). To do so, multiply by $A^+$ and realize $\hat{ \bv{x} } \equiv A^+ \bv{r}$ and $A^+ A = I$ to obtain

\[
\hat{ \bv{x} } - \bv{x} = A^+ \bv{\epsilon}
\]
the expected mean squared error is, by definition, the sum of the squared residuals,
\begin{align*}
    \mathcal{E}^2 =& \frac{1}{n} \left\langle 
         \sum_i \left( \hat{ x_i } - x_i \right)^2 
    \right\rangle \\
    =& \frac{1}{n}\left\langle 
        (A^+ \bv{\epsilon}) \cdot (A^+ \bv{\epsilon})
    \right\rangle \\
    =& \frac{\sigma^2}{n} \left\langle \mathrm{Tr} \left[ A^{+T} A^+ \right] \right\rangle \\
    =& \frac{\sigma^2}{n} \left\langle \mathrm{Tr} \left[ (A^T A)^{-1} \right] \right\rangle
\end{align*}
where in the last step we have expanded $A^+ = (A^T A)^{-1} A^T$ and used the circular property of the trace ($\mathrm{Tr} ABC = \mathrm{Tr} BCA$) twice.

\section{Traditional Ghost Imaging}
\label{sec:traditional_gi}

A quick remark is in order concerning the ``traditional'' algorithm for ghost imaging, as our discussion reveals how this method works and describes its performance. In the field of ghost imaging, the signal $\bv{x}$ is reconstructed via correlation with ``bucket'' detector measurements $r_i$ corresponding to many illumination patterns $\bv{a}_i$. Then the image estimate $\hat{\bv{x}}$ is obtained via correlation
\[
\hat{\bv{x}} - \langle x \rangle_i
= \Big\langle \bv{a}_i \cdot ( r_i - \langle r \rangle_i ) \Big\rangle_i
\]
in matrix notation
\begin{equation}
\label{eq:traditional_gi}
\hat{\bv{x}}  - \langle x \rangle_i 
= A^T ( \bv{r} - \langle r \rangle_i ) \,.
\end{equation}
However the forward model is still $\bv{r} = A \bv{x}$. Thus, the traditional ghost imaging algorithm implicitly assumes $A^TA \approx c_1 I + c_2 J$. The interesting observation is that while in general this assumption is not satisfied, we have shown that error-minimizing multiplexing schemes approach this behavior, explaining the success of the traditional ghost imaging algorithm.

That said, the traditional algorithm is inferior in both theory and practice to the ordinary least squares solution given by Eq.~(\ref{eq:ols-solution}). The Gauss-Markov theorem proves this so long as the noise is uncorrelated, zero-mean, and has identical variance for all measurement elements. 

Note: to obtain properly normalized estimates, like those presented in Fig.~\ref{fig:simulation}, one can use
\[
\hat{\bv{x}} = \frac{n A^T}{\mathrm{Tr} A^T A} \bv{r}
\]
where we have assumed $\hat{\bv{x}}$ and $\bv{r}$ are centered (mean-subtracted). If we let $\hat{\bv{x}} = c A^T \bv{r}$, this is the form for which
\[
c = \min_c || \hat{ \bv{x} } - \bv{x} ||_2^2
\]
Proof. Let $\delta \equiv (c A^T A - I) \bv{x}$. Then by definition,
\begin{align*}
    \epsilon =& \min_c \sum_j \left[ 
        \left( c A^T A - I \right) \bv{x}
    \right]^2_j
    \\
    =& \min_c \delta \cdot \delta 
\end{align*}
expand $\bv{x} = \sum_i b_i \phi_i$ in the basis of the eigenvectors $\phi_i$ of $G \equiv A^T A$ (a Grammian and therefore orthogonal matrix),
\begin{align*}
    \delta =& \sum_i b_i (cG - I) \phi_i \\
           =& \sum_i b_i (c \lambda_i - 1) \phi_i \\
           =& \, \bv{x} \cdot (c \bv{\lambda} - \bv{1} )
\end{align*}
now
\[
\delta \cdot \delta = (\bv{x} \cdot \bv{x}) 
\left[ (c \bv{\lambda} - \bv{1} ) \cdot (c \bv{\lambda} - \bv{1} ) \right]
\]
which is minimized for $c = n / \sum_i \lambda_i$.

\section{Additive Error}
\label{sec:additive_error}

The noise model we consider combines noise from three sources: unavoidable Poisson (quantum) noise, per-pixel Gaussian (Johnson type) noise, and finally per-measurement Gaussian (detector readout) noise. Each is shown to be well approximated by a Gaussian distribution. In this appendix we derive the expected variances of each of these types of noise individually; the final noise can be written as the sum of their variances. As a result we obtain that the error on the measurement $r_m = \bv{a}_m \cdot \bv{x}_m$ that is a mean-zero normal with variance
\[
\sigma^2 = \sigma_p^2 \sum_i a_{ni} + \sigma_x^2 \sum_i a_{ni}^2 
+ \sigma_m^2
\]
Here, $\sigma_p$ gives the magnitude of the Poisson noise relative to the per-pixel $\sigma_x$ and per-readout $\sigma_m^2$ Gaussian variances.

\subsection{Poisson Error}
\label{sec:poisson_error}
Consider summing photons recorded on a set of pixels, with each pixel's photon count distributed as $c_i \sim \mathrm{Pois}(a_{mi} \cdot x_{mi})$,
\[
r_m = \sum_i c_i \ \ \mathrm{for} \ \ c_i \sim \mathrm{Pois}(a_{mi} \cdot x_{mi})
\]
then $r_m \sim \mathrm{Pois}(\bv{a}_m \cdot \bv{x}_m)$, since the sum of iid Poissons is also a Poisson. The mean and variance of this distribution are both $\bv{a}_m \cdot \bv{x}_m$. For large values of this mean, above $\sim 1000$ (photons), a Gaussian approximation is very accurate, and $r_m \sim \mathcal{N}(\bv{a}_m \cdot \bv{x}_m, \bv{a}_m \cdot \bv{x}_m)$. We could equivalently write
\[
r_m = \bv{a}_m \cdot \bv{x}_m + \epsilon_m
\]
with $\epsilon_m \sim \mathcal{N}(0, \bv{a}_m \cdot \bv{x}_m)$. We wish, however, to consider the error for a generic measurement system independent of the structure of the sample under study. To do this we assume $\bv{x}_m \sim \sigma_p^2 \bv{1}$, with $\sigma_p^2$ a scale factor. Then the Poisson noise can be approximated by $\epsilon_m \sim \mathcal{N}(0, \sigma_p^2 \sum_i a_{mi})$.

\subsection{Per-Pixel Gaussian Error}
First, recall that for $\epsilon_i \stackrel{iid}{\sim} \mathcal{N}(0, \sigma_x^2)$ that $S \equiv \sum_i a_i \epsilon_i$ is distributed $S \sim \mathcal{N}(0, \sigma_x^2 \sum_i a_i^2)$. This means that if we have per-pixel errors $[\bv{\epsilon}_m]_i \stackrel{iid}{\sim} \mathcal{N}(0, \sigma_x^2)$ so that
\[
r_m = \bv{a}_m \cdot ( \bv{x}_m + \bv{\epsilon}_m )
\]
we can re-write those errors as a per-measurement error
\[
r_m = \bv{a}_m \cdot \bv{x}_m + \epsilon_m
\]
where $\epsilon_m \sim \mathcal{N}(0, \sigma_x^2 \sum_i a_{mi}^2)$.

\subsection{Per-Measurement Gaussian Error}
The simplest case is a per-measurement error, where 
\[
r_m = \bv{a}_m \cdot \bv{x}_m + \epsilon_m
\]
and $\epsilon_m \sim \mathcal{N}(0, \sigma_m^2)$ is a scalar additive error.

\section{IID Sensing Matrices}
\label{sec:iid_sensing}

Here we derive the error for sensing matrices with iid components. Let $A = (a_{ij})$ be such a matrix. Per Eq.~(\ref{eq:error}), our task is to evaluate the expected value
\[
\left\langle \mathrm{Tr} (A^T A)^{-1} \right\rangle
\]
Write the ($n \times n $) Grammian matrix $G = A^T A$ for convenience, the elements of which are
\[
g_{ij} = \sum_{k=1}^n a_{ki} a_{kj} = 
\begin{cases}
    m \langle a_{ii}^2 \rangle \ \ \mathrm{if} \ \ i=j \\
    m \langle a_{ij} \rangle^2 \ \ i \neq j
\end{cases}
\]
matrix $G$ has diagonal elements $d \equiv m \langle a_{ii}^2 \rangle$ and off-diagonal elements $c \equiv m \langle a_{ij} \rangle^2$. The eigenvalue equation $G \psi = \lambda \psi$ can be written
\begin{align*}
    \lambda \psi_1 &= d \psi_1 + n c \psi_2 \\
    \lambda \psi_2 &= d \psi_2 + c \psi_1 + (n-1) c \psi_2
\end{align*}
for all $n$ degenerate eigenvalues $\lambda$, which correspond to vectors $\psi$ with a single $\psi_1$ element and $n-1$ elements $\psi_2$. The solution to this system gives $\psi_2/\psi_1 = - 1/n$ and more relevantly
\[
\lambda = d-c = 
n \left( 
  \langle a_{ii}^2 \rangle - \langle a_{ij} \rangle^2 
\right)
\]
and so
\[
\left\langle \mathrm{Tr} G^{-1} \right\rangle = n \lambda^{-1}
=
\frac{n}{m}
\left( 
  \langle a_{ii}^2 \rangle - \langle a_{ij} \rangle^2 
\right)^{-1}
\]
the terms of this expression are real squared values and therefore positive. This shows that Grammian matrices that approximate the identity matrix, with large values on the diagonal and small off-diagonal matrices, have the lowest error, as expected.

For the half-Gaussian distribution with variance $a^2$, $\langle a_{ii}^2 \rangle = a^2$ and $\langle a_{ij} \rangle^2 = (2/\pi) a^2$ so $\left\langle \mathrm{Tr} G^{-1} \right\rangle = n / (m a^2) \cdot (1 - 2/\pi)^{-1}$.

For the Bernoulli distribution that gives value $a$ with probability $p$ and $0$ otherwise, $\langle a_{ii}^2 \rangle = p a^2$ and $\langle a_{ij} \rangle^2 = p^2 a^2$ so $\left\langle \mathrm{Tr} G^{-1} \right\rangle = (n/m) \left[ a^2 p(1-p) \right]^{-1}$.

\section{Generalized Linear Models}
\label{sec:generalizedmodels}

The loss function (\ref{eq:optimize}) is appropriate if the errors of the model are Gaussian. In some cases, however, this is not true and superior results can be obtained using a more appropriate model of the noise. One example that will be familiar to those working in x-ray or electron imaging is the low-flux situation, where error is primarily due to Poisson quantum noise, and the error is skewed as a result.

In such situations, it is possible to employ a generalized model to better model the error. Generalized Linear Models (GLMs) are well studied (as in \cite{Greene2003Econometric}). We only provide a brief overview here for completeness.

Instead of minimizing the least-squares error, as in (\ref{eq:optimize}), one can write a more general log-likelihood function
\[
\log \mathcal{L}(\bv{x}) = g(A \bv{x} | \bv{r}) + \gamma(\bv{x})
\]
Here, $g$ is a ``link function'' that provides a more general model to represent the likelihood of observing data $\bv{r}$ given input $A\bv{x}$, and $\gamma$ is a function capturing (all) prior beliefs that weights $\bv{x}$ according to their believed likelihood before any data are observed. 

The desired solution is then the maximum of the log-likelihood
\[
\hat{\bv{x}} = \mathrm{argmax}_{\bv{x}} \log \mathcal{L}(\bv{x})
\]
Equation (\ref{eq:optimize}) is a special case where $g$ is the Euclidean distance given by a Gaussian likelihood and no prior beliefs are asserted ($\gamma = \mathrm{const.}$).

Maximum-likelihood theory shows that the asymptotic distribution of $\hat{\bv{x}}$ is normally distributed around the true value $\bv{x}^*$, 
\[
\hat{\bv{x}} \sim \mathcal{N} ( \bv{x}^*, I(\bv{x}^*)^{-1} )
\]
with covariance given by the inverse of the Fisher Information
\[
I(\bv{x})_{ij} = - \left\langle
\frac{ \partial^2 \log \mathcal{L} (\bv{x}) }
{\partial \bv{x}_i \partial \bv{x}_j }
\right\rangle
\]
as previously, the expected mean squared error is
\[
    \mathcal{E}^2 = \frac{1}{n} \left\langle 
         \sum_i \left( \hat{ x_i } - x_i \right)^2 
    \right\rangle 
\]
but $\sum_i \left( \hat{ x_i } - x_i \right)^2$ is the trace of the covariance matrix, so
\[
   \mathcal{E}^2 = \frac{1}{n} \mathrm{Tr} \, I(\bv{x})^{-1}
\]
as a concrete and relevant example, consider the aforementioned Poisson regression case. Here, the link function is the exponential and the log likelihood becomes,
\[
\log \mathcal{L}(\bv{x}) = \sum_i \left[
- \exp( A_i \cdot \bv{x} ) + \bv{r}_i ( A_i \cdot \bv{x} ) - \log(\bv{r}_i !)
\right]
\]
where $A_i$ is the $i^{\mathrm{th}}$ row of $A$, corresponding to the $i^{\mathrm{th}}$ measurement. Taking the first derivative gives
\[
\frac{ \partial \log \mathcal{L} (\bv{x}) }
{\partial \bv{x} } = 
\sum_i \left[ \bv{r}_i - \exp( A_i \cdot \bv{x} ) \right] A_i
\]
which, when set to zero, gives a convex equation that can be solved numerically (via \textit{e.g.}~gradient ascent) to find the maximum likelihood solution. The second derivative is
\[
I(\bv{x}) = \sum_i A_i A_i^T \exp( A_i \cdot \bv{x} )
\]
One can in principle design an optimal sensing matrix $A$ by minimizing the trace of the inverse of this matrix. If we want to consider such a design before any knowledge of the signal $\bv{X}$ is know, we might approximate $A_i \cdot \bv{x} \approx \mathrm{const.}$ in which case we obtain the familiar 
\[
\mathcal{E}^2 = \mathrm{const.} \cdot \frac{1}{n}\mathrm{Tr} (A^T A)^{-1}
\]

\begin{acknowledgments}
TJL would like to thank Anders Nilsson for generously hosting him at Stockholm University during the time this paper was written. This work was supported by the operating contract for the Linac Coherent Light Source at SLAC National Accelerator Laboratory, DoE Contract No. DE-AC02-76SF00515
\end{acknowledgments}

\bibliography{use_of_gi}

\providecommand{\noopsort}[1]{}\providecommand{\singleletter}[1]{#1}%
\begin{thebibliography}{28}%
\makeatletter
\providecommand \@ifxundefined [1]{%
 \@ifx{#1\undefined}
}%
\providecommand \@ifnum [1]{%
 \ifnum #1\expandafter \@firstoftwo
 \else \expandafter \@secondoftwo
 \fi
}%
\providecommand \@ifx [1]{%
 \ifx #1\expandafter \@firstoftwo
 \else \expandafter \@secondoftwo
 \fi
}%
\providecommand \natexlab [1]{#1}%
\providecommand \enquote  [1]{``#1''}%
\providecommand \bibnamefont  [1]{#1}%
\providecommand \bibfnamefont [1]{#1}%
\providecommand \citenamefont [1]{#1}%
\providecommand \href@noop [0]{\@secondoftwo}%
\providecommand \href [0]{\begingroup \@sanitize@url \@href}%
\providecommand \@href[1]{\@@startlink{#1}\@@href}%
\providecommand \@@href[1]{\endgroup#1\@@endlink}%
\providecommand \@sanitize@url [0]{\catcode `\\12\catcode `\$12\catcode
  `\&12\catcode `\#12\catcode `\^12\catcode `\_12\catcode `\%12\relax}%
\providecommand \@@startlink[1]{}%
\providecommand \@@endlink[0]{}%
\providecommand \url  [0]{\begingroup\@sanitize@url \@url }%
\providecommand \@url [1]{\endgroup\@href {#1}{\urlprefix }}%
\providecommand \urlprefix  [0]{URL }%
\providecommand \Eprint [0]{\href }%
\providecommand \doibase [0]{https://doi.org/}%
\providecommand \selectlanguage [0]{\@gobble}%
\providecommand \bibinfo  [0]{\@secondoftwo}%
\providecommand \bibfield  [0]{\@secondoftwo}%
\providecommand \translation [1]{[#1]}%
\providecommand \BibitemOpen [0]{}%
\providecommand \bibitemStop [0]{}%
\providecommand \bibitemNoStop [0]{.\EOS\space}%
\providecommand \EOS [0]{\spacefactor3000\relax}%
\providecommand \BibitemShut  [1]{\csname bibitem#1\endcsname}%
\let\auto@bib@innerbib\@empty
\bibitem [{\citenamefont {Yorke}\ \emph {et~al.}(2014)\citenamefont {Yorke},
  \citenamefont {Beddard}, \citenamefont {Owen},\ and\ \citenamefont
  {Pearson}}]{Yorke:2014bda}%
  \BibitemOpen
  \bibfield  {author} {\bibinfo {author} {\bibfnamefont {B.~A.}\ \bibnamefont
  {Yorke}}, \bibinfo {author} {\bibfnamefont {G.~S.}\ \bibnamefont {Beddard}},
  \bibinfo {author} {\bibfnamefont {R.~L.}\ \bibnamefont {Owen}}, and\ \bibinfo
  {author} {\bibfnamefont {A.~R.}\ \bibnamefont {Pearson}},\ }\bibfield
  {title} {\bibinfo {title} {{Time-resolved crystallography using the Hadamard
  transform}},\ }\href@noop {} {\bibfield  {journal} {\bibinfo  {journal} {Nat
  Meth}\ }\textbf {\bibinfo {volume} {11}},\ \bibinfo {pages} {1131} (\bibinfo
  {year} {2014})}\BibitemShut {NoStop}%
\bibitem [{\citenamefont {Yu}\ \emph {et~al.}(2016)\citenamefont {Yu},
  \citenamefont {Lu}, \citenamefont {Han}, \citenamefont {Xie}, \citenamefont
  {Du}, \citenamefont {Xiao},\ and\ \citenamefont {Zhu}}]{Yu:2016hua}%
  \BibitemOpen
  \bibfield  {author} {\bibinfo {author} {\bibfnamefont {H.}~\bibnamefont
  {Yu}}, \bibinfo {author} {\bibfnamefont {R.}~\bibnamefont {Lu}}, \bibinfo
  {author} {\bibfnamefont {S.}~\bibnamefont {Han}}, \bibinfo {author}
  {\bibfnamefont {H.}~\bibnamefont {Xie}}, \bibinfo {author} {\bibfnamefont
  {G.}~\bibnamefont {Du}}, \bibinfo {author} {\bibfnamefont {T.}~\bibnamefont
  {Xiao}}, and\ \bibinfo {author} {\bibfnamefont {D.}~\bibnamefont {Zhu}},\
  }\bibfield  {title} {\bibinfo {title} {{Fourier-Transform Ghost Imaging with
  Hard X Rays}},\ }\href@noop {} {\bibfield  {journal} {\bibinfo  {journal}
  {Phys. Rev. Lett.}\ }\textbf {\bibinfo {volume} {117}},\ \bibinfo {pages}
  {259} (\bibinfo {year} {2016})}\BibitemShut {NoStop}%
\bibitem [{\citenamefont {Pelliccia}\ \emph {et~al.}(2016)\citenamefont
  {Pelliccia}, \citenamefont {Rack}, \citenamefont {Scheel}, \citenamefont
  {Cantelli},\ and\ \citenamefont {Paganin}}]{Pelliccia:2016dsa}%
  \BibitemOpen
  \bibfield  {author} {\bibinfo {author} {\bibfnamefont {D.}~\bibnamefont
  {Pelliccia}}, \bibinfo {author} {\bibfnamefont {A.}~\bibnamefont {Rack}},
  \bibinfo {author} {\bibfnamefont {M.}~\bibnamefont {Scheel}}, \bibinfo
  {author} {\bibfnamefont {V.}~\bibnamefont {Cantelli}}, and\ \bibinfo {author}
  {\bibfnamefont {D.~M.}\ \bibnamefont {Paganin}},\ }\bibfield  {title}
  {\bibinfo {title} {{Experimental X-Ray Ghost Imaging}},\ }\href@noop {}
  {\bibfield  {journal} {\bibinfo  {journal} {Phys. Rev. Lett.}\ }\textbf
  {\bibinfo {volume} {117}},\ \bibinfo {pages} {30} (\bibinfo {year}
  {2016})}\BibitemShut {NoStop}%
\bibitem [{\citenamefont {Zhang}\ \emph {et~al.}(2018)\citenamefont {Zhang},
  \citenamefont {He}, \citenamefont {Wu}, \citenamefont {Chen},\ and\
  \citenamefont {Wang}}]{Zhang:2018fl}%
  \BibitemOpen
  \bibfield  {author} {\bibinfo {author} {\bibfnamefont {A.-X.}\ \bibnamefont
  {Zhang}}, \bibinfo {author} {\bibfnamefont {Y.-H.}\ \bibnamefont {He}},
  \bibinfo {author} {\bibfnamefont {L.-A.}\ \bibnamefont {Wu}}, \bibinfo
  {author} {\bibfnamefont {L.-M.}\ \bibnamefont {Chen}}, and\ \bibinfo {author}
  {\bibfnamefont {B.-B.}\ \bibnamefont {Wang}},\ }\bibfield  {title} {\bibinfo
  {title} {{Tabletop x-ray ghost imaging with ultra-low radiation}},\
  }\href@noop {} {\bibfield  {journal} {\bibinfo  {journal} {Optica}\ }\textbf
  {\bibinfo {volume} {5}},\ \bibinfo {pages} {374} (\bibinfo {year}
  {2018})}\BibitemShut {NoStop}%
\bibitem [{\citenamefont {Pelliccia}\ \emph {et~al.}(2018)\citenamefont
  {Pelliccia}, \citenamefont {Olbinado}, \citenamefont {Rack}, \citenamefont
  {Kingston}, \citenamefont {Myers},\ and\ \citenamefont
  {Paganin}}]{Pelliccia:2018ff}%
  \BibitemOpen
  \bibfield  {author} {\bibinfo {author} {\bibfnamefont {D.}~\bibnamefont
  {Pelliccia}}, \bibinfo {author} {\bibfnamefont {M.~P.}\ \bibnamefont
  {Olbinado}}, \bibinfo {author} {\bibfnamefont {A.}~\bibnamefont {Rack}},
  \bibinfo {author} {\bibfnamefont {A.~M.}\ \bibnamefont {Kingston}}, \bibinfo
  {author} {\bibfnamefont {G.~R.}\ \bibnamefont {Myers}}, and\ \bibinfo
  {author} {\bibfnamefont {D.~M.}\ \bibnamefont {Paganin}},\ }\bibfield
  {title} {\bibinfo {title} {{Towards a practical implementation of X-ray ghost
  imaging with synchrotron light.}},\ }\href@noop {} {\bibfield  {journal}
  {\bibinfo  {journal} {IUCrJ}\ }\textbf {\bibinfo {volume} {5}},\ \bibinfo
  {pages} {428} (\bibinfo {year} {2018})}\BibitemShut {NoStop}%
\bibitem [{\citenamefont {Li}\ \emph {et~al.}(2018)\citenamefont {Li},
  \citenamefont {Cropp}, \citenamefont {Kabra}, \citenamefont {Lane},
  \citenamefont {Wetzstein}, \citenamefont {Musumeci},\ and\ \citenamefont
  {Ratner}}]{Li:2018ega}%
  \BibitemOpen
  \bibfield  {author} {\bibinfo {author} {\bibfnamefont {S.}~\bibnamefont
  {Li}}, \bibinfo {author} {\bibfnamefont {F.}~\bibnamefont {Cropp}}, \bibinfo
  {author} {\bibfnamefont {K.}~\bibnamefont {Kabra}}, \bibinfo {author}
  {\bibfnamefont {T.~J.}\ \bibnamefont {Lane}}, \bibinfo {author}
  {\bibfnamefont {G.}~\bibnamefont {Wetzstein}}, \bibinfo {author}
  {\bibfnamefont {P.}~\bibnamefont {Musumeci}}, and\ \bibinfo {author}
  {\bibfnamefont {D.}~\bibnamefont {Ratner}},\ }\bibfield  {title} {\bibinfo
  {title} {{Electron Ghost Imaging}},\ }\href@noop {} {\bibfield  {journal}
  {\bibinfo  {journal} {Phys. Rev. Lett.}\ }\textbf {\bibinfo {volume} {121}},\
  \bibinfo {pages} {114801} (\bibinfo {year} {2018})}\BibitemShut {NoStop}%
\bibitem [{\citenamefont {Kim}\ \emph {et~al.}(2018)\citenamefont {Kim},
  \citenamefont {Gelisio}, \citenamefont {Mercurio}, \citenamefont
  {Dziarzhytski}, \citenamefont {Beye}, \citenamefont {Bocklage}, \citenamefont
  {Classen}, \citenamefont {David}, \citenamefont {Gorobtsov}, \citenamefont
  {Khubbutdinov}, \citenamefont {Lazarev}, \citenamefont {Mukharamova},
  \citenamefont {Obukhov}, \citenamefont {oesner}, \citenamefont {Schlage},
  \citenamefont {Zaluzhnyy}, \citenamefont {Brenner}, \citenamefont
  {oehlsberger}, \citenamefont {von Zanthier}, \citenamefont {Wurth},\ and\
  \citenamefont {Vartanyants}}]{Kim:2018vu}%
  \BibitemOpen
  \bibfield  {author} {\bibinfo {author} {\bibfnamefont {Y.~Y.}\ \bibnamefont
  {Kim}}, \bibinfo {author} {\bibfnamefont {L.}~\bibnamefont {Gelisio}},
  \bibinfo {author} {\bibfnamefont {G.}~\bibnamefont {Mercurio}}, \bibinfo
  {author} {\bibfnamefont {S.}~\bibnamefont {Dziarzhytski}}, \bibinfo {author}
  {\bibfnamefont {M.}~\bibnamefont {Beye}}, \bibinfo {author} {\bibfnamefont
  {L.}~\bibnamefont {Bocklage}}, \bibinfo {author} {\bibfnamefont
  {A.}~\bibnamefont {Classen}}, \bibinfo {author} {\bibfnamefont
  {C.}~\bibnamefont {David}}, \bibinfo {author} {\bibfnamefont {O.~Y.}\
  \bibnamefont {Gorobtsov}}, \bibinfo {author} {\bibfnamefont {R.}~\bibnamefont
  {Khubbutdinov}}, \bibinfo {author} {\bibfnamefont {S.}~\bibnamefont
  {Lazarev}}, \bibinfo {author} {\bibfnamefont {N.}~\bibnamefont
  {Mukharamova}}, \bibinfo {author} {\bibfnamefont {Y.~N.}\ \bibnamefont
  {Obukhov}}, \bibinfo {author} {\bibfnamefont {B.~R.}\ \bibnamefont {oesner}},
  \bibinfo {author} {\bibfnamefont {K.}~\bibnamefont {Schlage}}, \bibinfo
  {author} {\bibfnamefont {I.~A.}\ \bibnamefont {Zaluzhnyy}}, \bibinfo {author}
  {\bibfnamefont {G.~u.}\ \bibnamefont {Brenner}}, \bibinfo {author}
  {\bibfnamefont {R.~R.}\ \bibnamefont {oehlsberger}}, \bibinfo {author}
  {\bibfnamefont {J.}~\bibnamefont {von Zanthier}}, \bibinfo {author}
  {\bibfnamefont {W.}~\bibnamefont {Wurth}}, and\ \bibinfo {author}
  {\bibfnamefont {I.~A.}\ \bibnamefont {Vartanyants}},\ }\bibfield  {title}
  {\bibinfo {title} {{Ghost Imaging at an XUV Free-Electron Laser}},\
  }\href@noop {} {\bibfield  {journal} {\bibinfo  {journal} {arXiv}\ }
  (\bibinfo {year} {2018})},\ \Eprint {https://arxiv.org/abs/1811.06855v1}
  {1811.06855v1} \BibitemShut {NoStop}%
\bibitem [{\citenamefont {Kingston}\ \emph {et~al.}(2018)\citenamefont
  {Kingston}, \citenamefont {Pelliccia}, \citenamefont {Rack}, \citenamefont
  {Olbinado}, \citenamefont {Cheng}, \citenamefont {Myers},\ and\ \citenamefont
  {Paganin}}]{Kingston:2018fl}%
  \BibitemOpen
  \bibfield  {author} {\bibinfo {author} {\bibfnamefont {A.~M.}\ \bibnamefont
  {Kingston}}, \bibinfo {author} {\bibfnamefont {D.}~\bibnamefont {Pelliccia}},
  \bibinfo {author} {\bibfnamefont {A.}~\bibnamefont {Rack}}, \bibinfo {author}
  {\bibfnamefont {M.~P.}\ \bibnamefont {Olbinado}}, \bibinfo {author}
  {\bibfnamefont {Y.}~\bibnamefont {Cheng}}, \bibinfo {author} {\bibfnamefont
  {G.~R.}\ \bibnamefont {Myers}}, and\ \bibinfo {author} {\bibfnamefont
  {D.~M.}\ \bibnamefont {Paganin}},\ }\bibfield  {title} {\bibinfo {title}
  {{Ghost tomography}},\ }\href@noop {} {\bibfield  {journal} {\bibinfo
  {journal} {Optica}\ }\textbf {\bibinfo {volume} {5}},\ \bibinfo {pages}
  {1516} (\bibinfo {year} {2018})}\BibitemShut {NoStop}%
\bibitem [{\citenamefont {Candes}\ and\ \citenamefont
  {Wakin}(2008)}]{Candes:2008hb}%
  \BibitemOpen
  \bibfield  {author} {\bibinfo {author} {\bibfnamefont {E.~J.}\ \bibnamefont
  {Candes}}and\ \bibinfo {author} {\bibfnamefont {M.~B.}\ \bibnamefont
  {Wakin}},\ }\bibfield  {title} {\bibinfo {title} {{An Introduction To
  Compressive Sampling}},\ }\href@noop {} {\bibfield  {journal} {\bibinfo
  {journal} {IEEE Signal Process. Mag.}\ }\textbf {\bibinfo {volume} {25}},\
  \bibinfo {pages} {21} (\bibinfo {year} {2008})}\BibitemShut {NoStop}%
\bibitem [{\citenamefont {Foucart}\ and\ \citenamefont
  {Rauhut}(2013)}]{Foucart:wp}%
  \BibitemOpen
  \bibfield  {author} {\bibinfo {author} {\bibfnamefont {S.}~\bibnamefont
  {Foucart}}and\ \bibinfo {author} {\bibfnamefont {H.}~\bibnamefont {Rauhut}},\
  }\href@noop {} {\emph {\bibinfo {title} {{A Mathematical Introduction to
  Compressive Sensing}}}}\ (\bibinfo  {publisher} {Springer},\ \bibinfo {year}
  {2013})\BibitemShut {NoStop}%
\bibitem [{\citenamefont {Padgett}\ and\ \citenamefont
  {Boyd}(2017)}]{Padgett:2017co}%
  \BibitemOpen
  \bibfield  {author} {\bibinfo {author} {\bibfnamefont {M.~J.}\ \bibnamefont
  {Padgett}}and\ \bibinfo {author} {\bibfnamefont {R.~W.}\ \bibnamefont
  {Boyd}},\ }\bibfield  {title} {\bibinfo {title} {{An introduction to ghost
  imaging: quantum and classical}},\ }\href@noop {} {\bibfield  {journal}
  {\bibinfo  {journal} {Philosophical Transactions of the Royal Society A:
  Mathematical, Physical and Engineering Sciences}\ }\textbf {\bibinfo {volume}
  {375}},\ \bibinfo {pages} {20160233} (\bibinfo {year} {2017})}\BibitemShut
  {NoStop}%
\bibitem [{\citenamefont {Shapiro}(2008)}]{Shapiro:2008dt}%
  \BibitemOpen
  \bibfield  {author} {\bibinfo {author} {\bibfnamefont {J.~H.}\ \bibnamefont
  {Shapiro}},\ }\bibfield  {title} {\bibinfo {title} {{Computational ghost
  imaging}},\ }\href@noop {} {\bibfield  {journal} {\bibinfo  {journal} {Phys.
  Rev. A}\ }\textbf {\bibinfo {volume} {78}},\ \bibinfo {pages} {061802}
  (\bibinfo {year} {2008})}\BibitemShut {NoStop}%
\bibitem [{\citenamefont {Sloane}(1979)}]{Sloane:1979tg}%
  \BibitemOpen
  \bibfield  {author} {\bibinfo {author} {\bibfnamefont {N.}~\bibnamefont
  {Sloane}},\ }\bibfield  {title} {\bibinfo {title} {{Multiplexing methods in
  spectroscopy}},\ }\href@noop {} {\bibfield  {journal} {\bibinfo  {journal}
  {Mathematics Magazine}\ }\textbf {\bibinfo {volume} {52}},\ \bibinfo {pages}
  {71} (\bibinfo {year} {1979})}\BibitemShut {NoStop}%
\bibitem [{\citenamefont {Harwit}\ and\ \citenamefont
  {Sloane}(1979)}]{Harwit:1979wf}%
  \BibitemOpen
  \bibfield  {author} {\bibinfo {author} {\bibfnamefont {M.}~\bibnamefont
  {Harwit}}and\ \bibinfo {author} {\bibfnamefont {N.}~\bibnamefont {Sloane}},\
  }\href@noop {} {\emph {\bibinfo {title} {{Hadamard Transform Optics}}}}\
  (\bibinfo  {publisher} {Academic Press},\ \bibinfo {year} {1979})\BibitemShut
  {NoStop}%
\bibitem [{Note1()}]{Note1}%
  \BibitemOpen
  \bibinfo {note} {We have not considered the case where there are errors in
  the record of the sensing matrix $A$. Such errors are possible when using
  random $A$ matrices generated naturally as part of the experiment and then
  measured by some diagnostic, for example as is done in ghost imaging (see
  Fig.~\ref {fig:sampling}). Sensing matrix errors make our estimates of $X$
  too small, an effect known as \protect \textit {regression dilution}.
  Regression dilution has been studied in detail in the field of statistics and
  we refer the interested reader to \cite {Draper:1998ab}.}\BibitemShut {Stop}%
\bibitem [{\citenamefont {Katz}\ \emph {et~al.}(2009)\citenamefont {Katz},
  \citenamefont {Bromberg},\ and\ \citenamefont {Silberberg}}]{Katz:2009fv}%
  \BibitemOpen
  \bibfield  {author} {\bibinfo {author} {\bibfnamefont {O.}~\bibnamefont
  {Katz}}, \bibinfo {author} {\bibfnamefont {Y.}~\bibnamefont {Bromberg}}, and\
  \bibinfo {author} {\bibfnamefont {Y.}~\bibnamefont {Silberberg}},\ }\bibfield
   {title} {\bibinfo {title} {{Compressive ghost imaging}},\ }\href@noop {}
  {\bibfield  {journal} {\bibinfo  {journal} {arXiv}\ ,\ \bibinfo {pages}
  {131110}} (\bibinfo {year} {2009})},\ \Eprint
  {https://arxiv.org/abs/0905.0321v2} {0905.0321v2} \BibitemShut {NoStop}%
\bibitem [{\citenamefont {Zhang}\ \emph {et~al.}(2014)\citenamefont {Zhang},
  \citenamefont {Guo}, \citenamefont {Cao}, \citenamefont {Guan},\ and\
  \citenamefont {Gao}}]{Zhang:2014go}%
  \BibitemOpen
  \bibfield  {author} {\bibinfo {author} {\bibfnamefont {C.}~\bibnamefont
  {Zhang}}, \bibinfo {author} {\bibfnamefont {S.}~\bibnamefont {Guo}}, \bibinfo
  {author} {\bibfnamefont {J.}~\bibnamefont {Cao}}, \bibinfo {author}
  {\bibfnamefont {J.}~\bibnamefont {Guan}}, and\ \bibinfo {author}
  {\bibfnamefont {F.}~\bibnamefont {Gao}},\ }\bibfield  {title} {\bibinfo
  {title} {{Object reconstitution using pseudo-inverse for ghost imaging}},\
  }\href@noop {} {\bibfield  {journal} {\bibinfo  {journal} {Opt Express}\
  }\textbf {\bibinfo {volume} {22}},\ \bibinfo {pages} {30063} (\bibinfo {year}
  {2014})}\BibitemShut {NoStop}%
\bibitem [{\citenamefont {Jiang}\ \emph {et~al.}(2019)\citenamefont {Jiang},
  \citenamefont {Li}, \citenamefont {Zhang}, \citenamefont {Jiang},
  \citenamefont {Wang}, \citenamefont {He}, \citenamefont {Wang},\ and\
  \citenamefont {Sun}}]{Jiang:19}%
  \BibitemOpen
  \bibfield  {author} {\bibinfo {author} {\bibfnamefont {S.}~\bibnamefont
  {Jiang}}, \bibinfo {author} {\bibfnamefont {X.}~\bibnamefont {Li}}, \bibinfo
  {author} {\bibfnamefont {Z.}~\bibnamefont {Zhang}}, \bibinfo {author}
  {\bibfnamefont {W.}~\bibnamefont {Jiang}}, \bibinfo {author} {\bibfnamefont
  {Y.}~\bibnamefont {Wang}}, \bibinfo {author} {\bibfnamefont {G.}~\bibnamefont
  {He}}, \bibinfo {author} {\bibfnamefont {Y.}~\bibnamefont {Wang}}, and\
  \bibinfo {author} {\bibfnamefont {B.}~\bibnamefont {Sun}},\ }\bibfield
  {title} {\bibinfo {title} {Scan efficiency of structured illumination in
  iterative single pixel imaging},\ }\href
  {https://doi.org/10.1364/OE.27.022499} {\bibfield  {journal} {\bibinfo
  {journal} {Opt. Express}\ }\textbf {\bibinfo {volume} {27}},\ \bibinfo
  {pages} {22499} (\bibinfo {year} {2019})}\BibitemShut {NoStop}%
\bibitem [{\citenamefont {F{\"o}rster}\ \emph {et~al.}(2019)\citenamefont
  {F{\"o}rster}, \citenamefont {Brandstetter},\ and\ \citenamefont
  {Schulze-Briese}}]{Forster:2019kz}%
  \BibitemOpen
  \bibfield  {author} {\bibinfo {author} {\bibfnamefont {A.}~\bibnamefont
  {F{\"o}rster}}, \bibinfo {author} {\bibfnamefont {S.}~\bibnamefont
  {Brandstetter}}, and\ \bibinfo {author} {\bibfnamefont {C.}~\bibnamefont
  {Schulze-Briese}},\ }\bibfield  {title} {\bibinfo {title} {{Transforming
  X-ray detection with hybrid photon counting detectors.}},\ }\href@noop {}
  {\bibfield  {journal} {\bibinfo  {journal} {Philos Trans A Math Phys Eng
  Sci}\ }\textbf {\bibinfo {volume} {377}},\ \bibinfo {pages} {20180241}
  (\bibinfo {year} {2019})}\BibitemShut {NoStop}%
\bibitem [{\citenamefont {K{\"u}hlbrandt}(2014)}]{Kuhlbrandt:2014ho}%
  \BibitemOpen
  \bibfield  {author} {\bibinfo {author} {\bibfnamefont {W.}~\bibnamefont
  {K{\"u}hlbrandt}},\ }\bibfield  {title} {\bibinfo {title} {{Biochemistry. The
  resolution revolution.}},\ }\href@noop {} {\bibfield  {journal} {\bibinfo
  {journal} {Science}\ }\textbf {\bibinfo {volume} {343}},\ \bibinfo {pages}
  {1443} (\bibinfo {year} {2014})}\BibitemShut {NoStop}%
\bibitem [{Note2()}]{Note2}%
  \BibitemOpen
  \bibinfo {note} {These matrices are closely related to, but are not exactly,
  the traditionally defined Hadamard matrices. They are formed by generating a
  Hadamard matrix of order $n+1$, truncating the first row and column -- which
  contain only 1s -- and performing the substitions $1 \to 0$ and $-1 \to
  +1$.}\BibitemShut {Stop}%
\bibitem [{\citenamefont {Behrens}\ \emph {et~al.}(2014)\citenamefont
  {Behrens}, \citenamefont {Decker}, \citenamefont {Ding}, \citenamefont
  {Dolgashev}, \citenamefont {Frisch}, \citenamefont {Huang}, \citenamefont
  {Krejcik}, \citenamefont {Loos}, \citenamefont {Lutman}, \citenamefont
  {Maxwell}, \citenamefont {Turner}, \citenamefont {Wang}, \citenamefont
  {Wang}, \citenamefont {Welch},\ and\ \citenamefont {Wu}}]{Behrens:2014ib}%
  \BibitemOpen
  \bibfield  {author} {\bibinfo {author} {\bibfnamefont {C.}~\bibnamefont
  {Behrens}}, \bibinfo {author} {\bibfnamefont {F.~J.}\ \bibnamefont {Decker}},
  \bibinfo {author} {\bibfnamefont {Y.}~\bibnamefont {Ding}}, \bibinfo {author}
  {\bibfnamefont {V.~A.}\ \bibnamefont {Dolgashev}}, \bibinfo {author}
  {\bibfnamefont {J.}~\bibnamefont {Frisch}}, \bibinfo {author} {\bibfnamefont
  {Z.}~\bibnamefont {Huang}}, \bibinfo {author} {\bibfnamefont
  {P.}~\bibnamefont {Krejcik}}, \bibinfo {author} {\bibfnamefont
  {H.}~\bibnamefont {Loos}}, \bibinfo {author} {\bibfnamefont {A.}~\bibnamefont
  {Lutman}}, \bibinfo {author} {\bibfnamefont {T.~J.}\ \bibnamefont {Maxwell}},
  \bibinfo {author} {\bibfnamefont {J.}~\bibnamefont {Turner}}, \bibinfo
  {author} {\bibfnamefont {J.}~\bibnamefont {Wang}}, \bibinfo {author}
  {\bibfnamefont {M.~H.}\ \bibnamefont {Wang}}, \bibinfo {author}
  {\bibfnamefont {J.}~\bibnamefont {Welch}}, and\ \bibinfo {author}
  {\bibfnamefont {J.}~\bibnamefont {Wu}},\ }\bibfield  {title} {\bibinfo
  {title} {{Few-femtosecond time-resolved measurements of X-ray free-electron
  lasers.}},\ }\href@noop {} {\bibfield  {journal} {\bibinfo  {journal} {Nature
  Communications}\ }\textbf {\bibinfo {volume} {5}},\ \bibinfo {pages} {3762}
  (\bibinfo {year} {2014})}\BibitemShut {NoStop}%
\bibitem [{\citenamefont {MacArthur}\ \emph {et~al.}(2017)\citenamefont
  {MacArthur}, \citenamefont {Duris}, \citenamefont {Huang},\ and\
  \citenamefont {Marinelli}}]{MacArthur:IPAC2017-WEPAB118}%
  \BibitemOpen
  \bibfield  {author} {\bibinfo {author} {\bibfnamefont {J.}~\bibnamefont
  {MacArthur}}, \bibinfo {author} {\bibfnamefont {J.}~\bibnamefont {Duris}},
  \bibinfo {author} {\bibfnamefont {Z.}~\bibnamefont {Huang}}, and\ \bibinfo
  {author} {\bibfnamefont {A.}~\bibnamefont {Marinelli}},\ }\bibfield  {title}
  {\bibinfo {title} {High power sub-femtosecond x-ray pulse study for the
  lcls},\ }in\ \href
  {https://doi.org/https://doi.org/10.18429/JACoW-IPAC2017-WEPAB118} {\emph
  {\bibinfo {booktitle} {Proc. of International Particle Accelerator Conference
  (IPAC'17), Copenhagen, Denmark, 14-19 May, 2017}}},\ \bibinfo {series and
  number} {\bibinfo {series} {International Particle Accelerator Conference}\
  No.~\bibinfo {number} {8}}\ (\bibinfo  {publisher} {JACoW},\ \bibinfo
  {address} {Geneva, Switzerland},\ \bibinfo {year} {2017})\ pp.\ \bibinfo
  {pages} {2848--2850},\ \bibinfo {note}
  {https://doi.org/10.18429/JACoW-IPAC2017-WEPAB118}\BibitemShut {NoStop}%
\bibitem [{\citenamefont {Hartmann}\ \emph {et~al.}(2018)\citenamefont
  {Hartmann}, \citenamefont {Hartmann}, \citenamefont {Heider}, \citenamefont
  {Wagner}, \citenamefont {Ilchen}, \citenamefont {Buck}, \citenamefont
  {Lindahl}, \citenamefont {Benko}, \citenamefont {Gr{\"u}nert}, \citenamefont
  {Krzywinski}, \citenamefont {Liu}, \citenamefont {Lutman}, \citenamefont
  {Marinelli}, \citenamefont {Maxwell}, \citenamefont {Miahnahri},
  \citenamefont {Moeller}, \citenamefont {Planas}, \citenamefont {Robinson},
  \citenamefont {Kazansky}, \citenamefont {Kabachnik}, \citenamefont
  {Viefhaus}, \citenamefont {Feurer}, \citenamefont {Kienberger}, \citenamefont
  {Coffee},\ and\ \citenamefont {Helml}}]{Hartmann:2018dq}%
  \BibitemOpen
  \bibfield  {author} {\bibinfo {author} {\bibfnamefont {N.}~\bibnamefont
  {Hartmann}}, \bibinfo {author} {\bibfnamefont {G.}~\bibnamefont {Hartmann}},
  \bibinfo {author} {\bibfnamefont {R.}~\bibnamefont {Heider}}, \bibinfo
  {author} {\bibfnamefont {M.~S.}\ \bibnamefont {Wagner}}, \bibinfo {author}
  {\bibfnamefont {M.}~\bibnamefont {Ilchen}}, \bibinfo {author} {\bibfnamefont
  {J.}~\bibnamefont {Buck}}, \bibinfo {author} {\bibfnamefont {A.~O.}\
  \bibnamefont {Lindahl}}, \bibinfo {author} {\bibfnamefont {C.}~\bibnamefont
  {Benko}}, \bibinfo {author} {\bibfnamefont {J.}~\bibnamefont {Gr{\"u}nert}},
  \bibinfo {author} {\bibfnamefont {J.}~\bibnamefont {Krzywinski}}, \bibinfo
  {author} {\bibfnamefont {J.}~\bibnamefont {Liu}}, \bibinfo {author}
  {\bibfnamefont {A.~A.}\ \bibnamefont {Lutman}}, \bibinfo {author}
  {\bibfnamefont {A.}~\bibnamefont {Marinelli}}, \bibinfo {author}
  {\bibfnamefont {T.}~\bibnamefont {Maxwell}}, \bibinfo {author} {\bibfnamefont
  {A.~A.}\ \bibnamefont {Miahnahri}}, \bibinfo {author} {\bibfnamefont {S.~P.}\
  \bibnamefont {Moeller}}, \bibinfo {author} {\bibfnamefont {M.}~\bibnamefont
  {Planas}}, \bibinfo {author} {\bibfnamefont {J.}~\bibnamefont {Robinson}},
  \bibinfo {author} {\bibfnamefont {A.~K.}\ \bibnamefont {Kazansky}}, \bibinfo
  {author} {\bibfnamefont {N.~M.}\ \bibnamefont {Kabachnik}}, \bibinfo {author}
  {\bibfnamefont {J.}~\bibnamefont {Viefhaus}}, \bibinfo {author}
  {\bibfnamefont {T.}~\bibnamefont {Feurer}}, \bibinfo {author} {\bibfnamefont
  {R.}~\bibnamefont {Kienberger}}, \bibinfo {author} {\bibfnamefont {R.~N.}\
  \bibnamefont {Coffee}}, and\ \bibinfo {author} {\bibfnamefont
  {W.}~\bibnamefont {Helml}},\ }\bibfield  {title} {\bibinfo {title}
  {{Attosecond time{\textendash}energy structure of X-ray free- electron laser
  pulses}},\ }\href@noop {} {\bibfield  {journal} {\bibinfo  {journal} {Nature
  Photon}\ ,\ \bibinfo {pages} {1}} (\bibinfo {year} {2018})}\BibitemShut
  {NoStop}%
\bibitem [{\citenamefont {Serkez}\ \emph {et~al.}(2018)\citenamefont {Serkez},
  \citenamefont {Geloni}, \citenamefont {Tomin}, \citenamefont {Feng},
  \citenamefont {Gryzlova}, \citenamefont {Grum-Grzhimailo},\ and\
  \citenamefont {Meyer}}]{Serkez:2018jx}%
  \BibitemOpen
  \bibfield  {author} {\bibinfo {author} {\bibfnamefont {S.}~\bibnamefont
  {Serkez}}, \bibinfo {author} {\bibfnamefont {G.}~\bibnamefont {Geloni}},
  \bibinfo {author} {\bibfnamefont {S.}~\bibnamefont {Tomin}}, \bibinfo
  {author} {\bibfnamefont {G.}~\bibnamefont {Feng}}, \bibinfo {author}
  {\bibfnamefont {E.~V.}\ \bibnamefont {Gryzlova}}, \bibinfo {author}
  {\bibfnamefont {A.~N.}\ \bibnamefont {Grum-Grzhimailo}}, and\ \bibinfo
  {author} {\bibfnamefont {M.}~\bibnamefont {Meyer}},\ }\bibfield  {title}
  {\bibinfo {title} {{Overview of options for generating high-brightness
  attosecond x-ray pulses at free-electron lasers and applications at the
  European XFEL}},\ }\href@noop {} {\bibfield  {journal} {\bibinfo  {journal}
  {J. Opt.}\ }\textbf {\bibinfo {volume} {20}},\ \bibinfo {pages} {024005}
  (\bibinfo {year} {2018})}\BibitemShut {NoStop}%
\bibitem [{\citenamefont {Draper}\ and\ \citenamefont
  {Smith}(1998)}]{Draper:1998ab}%
  \BibitemOpen
  \bibfield  {author} {\bibinfo {author} {\bibfnamefont {N.}~\bibnamefont
  {Draper}}and\ \bibinfo {author} {\bibfnamefont {H.}~\bibnamefont {Smith}},\
  }\href@noop {} {\emph {\bibinfo {title} {{Applied Regression Analysis}}}}\
  (\bibinfo  {publisher} {John Wiley},\ \bibinfo {year} {1998})\BibitemShut
  {NoStop}%
\bibitem [{Note3()}]{Note3}%
  \BibitemOpen
  \bibinfo {note} {We are assuming a complete column of $k$ elements in $X_s$
  is zero or not for the purposes of counting sparse elements. If one element
  is non-zero, that entire column is counted as one of the $s$ datapoints we
  have to infer.}\BibitemShut {Stop}%
\bibitem [{\citenamefont {Greene}(2003)}]{Greene2003Econometric}%
  \BibitemOpen
  \bibfield  {author} {\bibinfo {author} {\bibfnamefont {W.~H.}\ \bibnamefont
  {Greene}},\ }\href
  {http://pages.stern.nyu.edu/~wgreene/Text/econometricanalysis.htm} {\emph
  {\bibinfo {title} {Econometric Analysis}}},\ \bibinfo {edition} {5th}\ ed.\
  (\bibinfo  {publisher} {Pearson Education},\ \bibinfo {year}
  {2003})\BibitemShut {NoStop}%
\end{thebibliography}%

\end{document}